\documentclass[twocolumn,prx,aps,amsmath,amssymb,superscriptaddress,floatfix]{revtex4}

\usepackage{graphicx}
\usepackage{epstopdf}
\usepackage{epsfig}
\usepackage{amsbsy}
\usepackage{color}
\usepackage{mathtools}
\usepackage{cancel}
\usepackage{multirow}
\usepackage[dvipsnames]{xcolor}

\DeclarePairedDelimiterX\braket[2]{\langle}{\rangle}{#1 \delimsize\vert #2}

\newcommand{\reff}[1]{(\ref{#1})}
\newcommand{\fig}[1]{Fig. \ref{#1}}

\begin{document}
\title {Double quantum dots defined in bilayer graphene}

\author{D.P. \.Zebrowski}
 \affiliation{AGH University of Science and Technology, \\ Faculty of Physics and Applied Computer Science,
al. Mickiewicza 30, 30-059 Krak\'ow, Poland}

\author{F.M. Peeters}
 \affiliation{Departement Fysica, Universiteit Antwerpen,\\  Groenenborgerlaan 171, B-2020 Antwerpen, Belgium}

\author{B. Szafran}
 \affiliation{AGH University of Science and Technology, \\ Faculty of Physics and Applied Computer Science,
al. Mickiewicza 30, 30-059 Krak\'ow, Poland}

\date{\today}

\begin{abstract}
Artificial molecular states of double quantum dots defined  in bilayer graphene  are studied with the atomistic tight-binding and its low-energy continuum approximation. 
We indicate that the extended electron wave functions have opposite parities  on sublattices of the layers and that the ground-state 
 wave function components change from  bonding to  antibonding  with  the interdot distance. 
In the weak coupling limit -- the most relevant for the quantum dots defined electrostatically -- 
the signatures of the interdot coupling include -- for the two-electron ground state -- formation of states with symmetric or antisymmetric spatial wave functions  split  by the exchange energy.
In the high energy part of the spectrum the states with both electrons in the same dot are found with the splitting of energy levels corresponding to simultaneous tunneling of the electron pair from one dot to the other.

\end{abstract}

\maketitle

\section{Introduction}

Studies of artificial molecules formed by carrier orbitals extended over double quantum dots (DQDs)
focus on carrier tunneling and interactions  \cite{vqds,vqds1,ab1,ab2,ab3,paulisb}. A particular attention is payed to the spin-related
phenomena in the context of the quantum information processing \cite{ldp} using exchange
interaction due to the interdot coupling \cite{burkard,paulisb}. The coupling with nuclear spins \cite{decoh1,decoh2,decoh3,decoh4}
which limits the electron spin coherence times motivated studies of DQDs with holes as spin carriers \cite{douho,bohu}
as well as on systems based on silicon \cite{sili} and carbon, including nanotubes \cite{laird} 
and graphene \cite{tr}. In graphene the carrier storage is hampered by the Klein tunneling \cite{klein} but
for nanoribbons the lateral confinement opens the transport gap \cite{ribbons-gap} that makes the carrier storage 
possible \cite{dqdsingraphene}. Quantum dots in ribbons are influenced by edge effects and disorder \cite{nrb5}. 
An alternative medium is the bilayer graphene \cite{review} for which the perpendicular electric field opens the band gap \cite{gap} in the energy spectrum
and allows for carrier confinement by lateral fields \cite{biqd0,biqd1,biqd2,biqd3,biqd4,bizebr}. 

In this work we consider the bilayer graphene \cite{review} and formation of  extended  orbitals within the DQDs. In the low-energy continuum approach Hamiltonian eigenstates possess a definite  total angular momentum \cite{biqd0}
with a different orbital angular momentum for each of the four sublattices  described by the wave function components.
 % which corresponds to a various orbital momementum quantum number for each of the components. 
As we discuss below for a lower -- point symmetry of the DQD system -- the Hamiltonian eigenstates components correspond to opposite spatial parities for each of the sublattices. Thus wave function can be bonding 
on one sublattice and  antibonding on the other. The effect induces a complex dependence
of the spectrum on the DQD distance that is similar
to the antibonding heavy hole ground state found  for vertical self-assembled quantum dots \cite{ab1,ab2,ab3,rev1}. 

In the low energy part of the two-electron spectrum for the weak coupling case - the typical one for the electrostatic quantum dots 
-- the electrons are localized in separate QDs
and the ground state at zero magnetic field $B$ is nearly 16-times degenerate with respect to the valley and the spin.
For nonzero magnetic field the energy levels correspond to wave function that can be approximately described 
by a product of the spatial, valley and spin components. Only the symmetry of the spatial part against the electron interchange
 influences the energy of the states, and we find that the energy levels shift in pairs with $B$ that correspond to the opposite symmetry split by the exchange integral.  In the high energy 
part of the spectrum the states corresponding to both electrons in the same dot are found. These energy levels also shift in  $B$ in pairs which is a manifestation of a collective two-electron intedot tunneling  that forms bonding and antibonding two-electron orbitals. For stronger interdot coupling the DQD spectrum resembles the one of a single
quantum dot  with a  ground state triplet  \cite{zarenia2e}.

This paper is organized as follows. In the Theory Section (II) we describe the model system, the atomistic and continuum approaches, as well as
the configuration interaction method for the electron pair. In  Section  III we  discuss formation of the extended orbitals and their nature as artificial molecular states
and next we discuss the spatial, spin and orbital symmetries of the two-electron states. Summary and conclusions are given in Section IV.

\section{Theory}
\subsection{Model structure}
\begin{figure}[!h]
\includegraphics[scale=0.2 ]{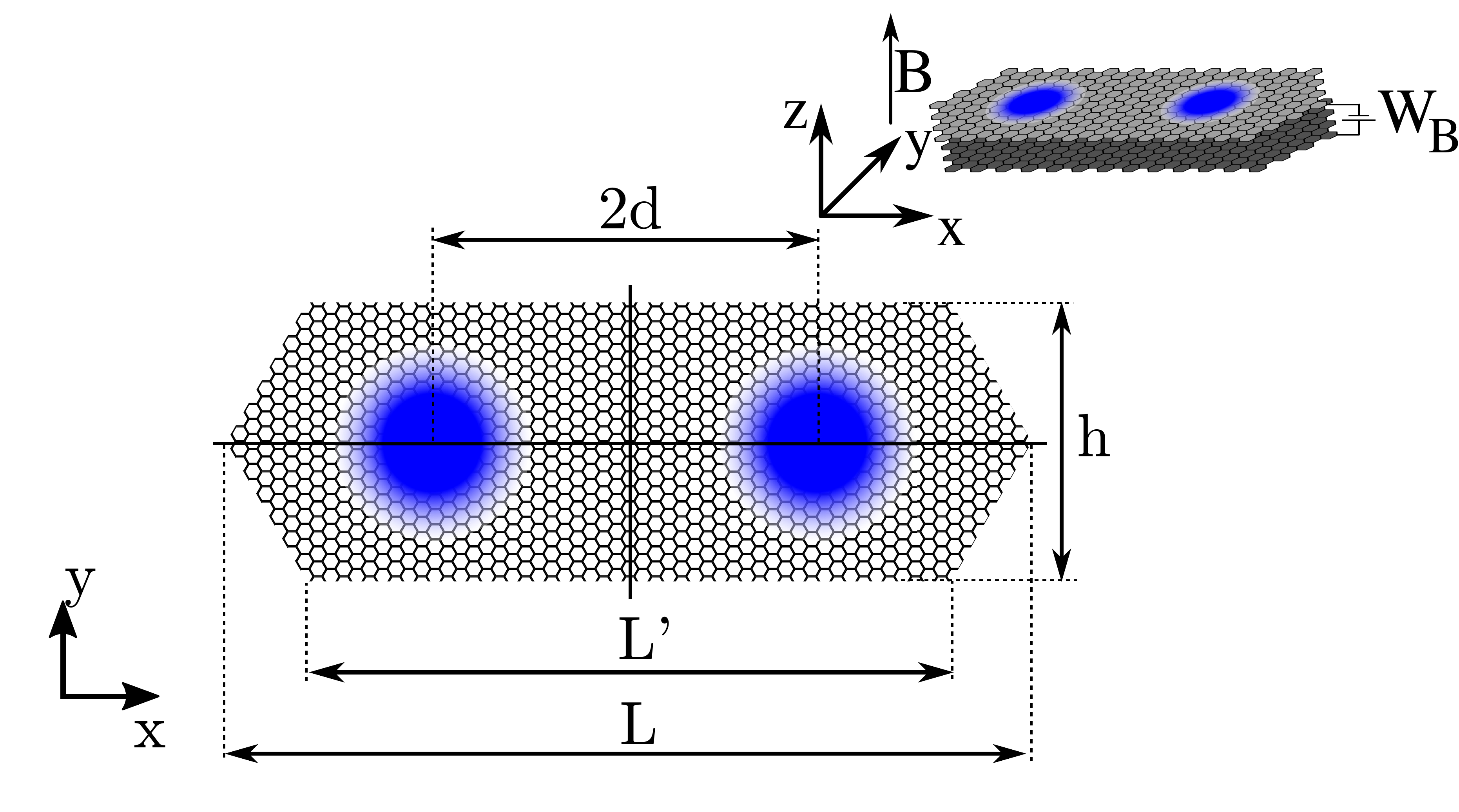}

 \caption{Schematic drawing of the considered system. Bilayer graphene flake in a hexagonal shape stretched in the x direction with quantum dots separated by $2d$ represented by blue circles.
We consider the flake of two sizes.
 The length of the smaller system is $L=64.6$ nm, the height is $h=27.6$ nm and the longer edge has length of $L'=48.8$ nm,
while the parameters for the larger are $L=89.46$ nm, $L'=63.9 $ nm and $h=44.51$ nm.}
 \label{uklad1}
\end{figure}

We consider a flake in form of a stretched hexagon (see \fig{uklad1})
with the bilayers in Bernal stacking \cite{review}
and two quantum dots defined by external potential.
We consider two sizes of the flake. 
The smaller flake has a length of  $L=64.6$ nm and height $h=27.6$ nm, with the horizontal edge length of $L'=48.8$ nm,   and the total number of carbon atoms $119928$.
The larger flake is characterized by $L=89.46$ nm, $L'=63.9 $ nm, $h=44.51$ nm, and contains $261000$ atoms. 
Two different sizes of the flake for fixed external potential defining the quantum dots are introduced for the discussion of the effects of the coupling between the dot-confined states and the edge of the flake.

We consider a flake with an armchair boundary for which no edge-localized energy levels appear in the spectrum at the neutrality point \cite{HexTriang,bizebr,zeb2}. 
The quantum dots for electrons of the conduction band are defined by external potential which is taken in form
\begin{equation}
 V_{QD} (x,y)=-V\exp(-((|x|-d)^2+y^2)/R^2),
 \label{qdpot}
\end{equation}
where the origin ($x=y=0$) is placed in the center of the flake,
 $V$ is the quantum dots depth, $R$ is the effective radius and $2d$ is distance between the centers of the QDs.  
In this work, we fix $R$ at $4$ nm, for this value the wave functions localized in the dots vanish off the edges of the larger flake -
for which the valley mixing effect of the armchair edge is negligible. For the smaller flake the effect of the edge for the confined states is still present
and we use this fact below for description of the valley mixing effects on the single- and the two-electron spectra.

\subsection{Tight-binding model}
For most of the calculations we use the atomistic tight-binding Hamiltonian \cite{review} given by
\begin{equation}
  \hat{H}_{1e}=\sum_{i,\sigma}W^{\sigma}_i\hat{c}^{\dagger}_{i,\sigma}\hat{c}_{i,\sigma}+\sum_{ ij,\sigma\sigma'} \left( \hat{c}^{\dagger}_{i,\sigma}\hat{c}_{j,\sigma'}t_{ij}+h.c \right ),
  \label{Hone}
\end{equation}
where $\hat{c}_{i,\sigma}^{\dagger}$ ( $\hat{c}_{i,\sigma}$)  creates (annihilates) an electron with spin $\sigma$ at ion $i$ and $t_{ij}$ is the hopping parameter.
At zero magnetic field the hopping parameter takes the value of $t_{ij}=-2.6$ eV for the in-plane nearest neighbor atoms and $t_{ij}=0.3$ eV for the vertical dimers
between the layers \cite{review}. 
In order to account for the  magnetic field perpendicular to the layers we include the Peierls phase into the hopping parameters,
\begin{equation}
t_{ij} \rightarrow t_{ij}e^{i\frac{e}{\hbar}\int_{\mathbf{r_i}}^{\mathbf{r_j}}\mathbf{A}\cdot \mathbf{dl}}, 
\end{equation}
where $\mathbf{B}=\mathbf{\nabla} \times \mathbf{A}$ with $\mathbf{A}=(0,B_zx,0)$.
The potential $W^{\sigma}_i$ in Eq. \reff{Hone} has the form
\begin{equation}
 W^{\sigma}(x,y)=V_{QD} (x,y)+\frac{W_{B}}{2}\mathbf{ \tau}_z+\frac{1}{2}\mu_Bg\sigma_z B_z, \label{pu}
\end{equation}
where the first term is given by Eq. ({\ref{qdpot}}) and the second one is the potential difference between the layers with $ \tau_z=\pm 1$, with $+$ ($-$) for the upper (lower) layer. This difference opens the energy gap in the bilayer flake by the  asymmetry introduced between the layers \cite{review}.
In our calculations we set $W_{B}=300$ meV which is equivalent to applying the vertical electric field 
 $F\approx 0.9$ V/nm. For this value we obtain the  gap of about $E_g\approx141$ meV. The last term in Eq. (\ref{pu}) introduces the Zeeman interaction, where $g=2$ is the Land\'e factor and $\sigma_z$ stands for the Pauli matrix.

\subsection{Continuum approximation}

For analysis of the single-electron envelope wave functions, we also consider the low-energy approximation in the continuum model. 
Near a single valley the electron wave functions are written in form of the four component spinor \cite{review}
\begin{equation}
 \Psi(\mathbf{r})=(\phi_A(\mathbf{r}),\phi_B(\mathbf{r}),\phi_{B'}(\mathbf{r}),\phi_{A'}(\mathbf{r}))^T,
\end{equation}
where $A,B$ ($B',A'$) represents the $A$ and $B$ sublattices of the upper (lower) layer respectively.
The Dirac Hamiltonian \cite{review}  around the $\mathbf{K}$ valley takes the form
\begin{equation}
\hat{H}_{D}=\begin{pmatrix} V_{QD}& \pi & t_\bot & 0 \\ \pi^{\dagger} & V_{QD}& 0 & 0 \\ t_\bot & 0 & V_{QD}& \pi^{\dagger}\\ 0 & 0 & \pi & V_{QD}\end{pmatrix}+\frac{W_{B}}{2}\mathbf{\tau}_z,
\label{HDirac}
\end{equation}
where $\pi=v_F\left(p_x+i p_y \right)$, $(p_x,p_y)$ is  the momentum operator, $v_F=\frac{3ta}{2}=0.84 \cdot 10^6 \frac{m}{s}$ -- the Fermi velocity,
 and $t_\bot=0.3$ eV is the interlayer coupling term. 
The $\tau_z$ operator defined as 
\begin{equation}
 \tau_z=\begin{pmatrix} \mathbb{I}_2 & 0 \\ 0 & -\mathbb{I}_2\end{pmatrix},
 \label{layeri}
\end{equation}
that assigns +1 (-1) for upper (lower) layer respectively opens the gap by applying the voltage difference ($W_{B}$) between the layers.

The quantum dots potential \reff{qdpot} is symmetric with respect to the point inversion at the origin.
For this potential the Hamiltonian commutes with a generalized parity operator 
\begin{equation}
 \hat{U}_P=\begin{pmatrix} -\sigma_z & 0 \\ 0 & -\sigma_z\end{pmatrix}\hat{P},
\end{equation}
where $\hat{P}$ changes the sign of the spatial coordinates of scalar functions $\hat{P}\phi(\mathbf{r})=\pm\phi(\mathbf{-r}) $,
 and $\mathbb{I}_n$ is the $n\times n $ identity matrix.

For a single circular quantum dot the total angular momentum operator 
\begin{equation}
 \hat{J}_z=\hat{L}_z\mathbb{I}_4+\frac{\hbar}{2}  \tau_z  -\frac{\hbar}{2}\begin{pmatrix} \sigma_z & 0 \\ 0 & -\sigma_z\end{pmatrix}   ,
 \label{angmom}
\end{equation}
commutes with the Hamiltonian \cite{biqd0}, 
where $\hat{L}_z$ is the angular momentum operator. The second term  is the layer index operator with $\tau_z$ defined by \reff{layeri}. The last term corresponds to the pseudospin \cite{review}. 

In order to find the eigenstates of the Hamiltonian \reff{HDirac} we use the finite element method with triangular elements and  cubic shape functions.
The fermion doubling  fast-varying spurious solutions \cite{Wilson} are eliminated  
with the additional  Wilson \cite{Wilson} term introduced to the Hamiltonian \reff{HDirac} 
\begin{equation}
 \hat{H}_W=W_P \hbar v_F a_D  \begin{pmatrix} -\sigma_z & 0 \\ 0 & -\sigma_z\end{pmatrix}\nabla^2,
\end{equation}
where $W_P=0.015$ is the dimensionless Wilson parameter \cite{Wilson} and $a_D=1.32$ nm is the discretization constant. $W_P$ parameter
was fine-tuned to remove  the spurious solutions from the discussed energy range leaving 
 the actual states  nearly unaffected.

\subsection{Configuration-interaction method}

 The wave functions obtained with the tight-binding approach are used in the configuration interaction method \cite{tbci,tbci2}.
The atomistic approach when applied to the exact diagonalization method naturally accounts for the intervalley scattering induced by the short range
component of the Coulomb interaction \cite{vm1,vm2,vm3}.

For the description of the two-electron states we use the solution of the one-electron eigenproblem \reff{Hone}. % for $B=0$.
Then, we expand the two-electron wave function in the basis of $M$ Slater determinants
\begin{equation}
 \Psi=\sum_{i=1}^{M} d_{i}\left( \psi_{i1}(\mathbf{x_1})\otimes \psi_{i2}(\mathbf{x_2})-\psi_{i2}(\mathbf{x_1})\otimes \psi_{i1}(\mathbf{x_2})  \right) 
 \label{twoele}
\end{equation}
with spin-orbitals $\psi_{i}(\mathbf{x}),\ i=1,2,\dots M$ where $\mathbf{x}=(\mathbf{r},\sigma)$ represents the orbital and spin coordinates. 
The number of Slater determinants $M$ is ${K}\choose{2} $, where $K$ is the number of dot-localized single-electron spin-orbitals. 
 The two-electron Hamiltonian is 
\begin{equation}
  \hat{H}_{2e}(\mathbf{x_1},\mathbf{x_2})=\sum_{i=1}^2\hat{H}_{1e}(\mathbf{x_i})+\frac{\kappa}{|\mathbf{r_{12}}|},
\end{equation}
where  $\kappa=e^2/(4\pi\epsilon\epsilon_0)$. The dielectric constant $\epsilon=6$ is taken from Ref. \cite{sicc} for graphene grown on SiC.
Using, the form of wave function given by \reff{twoele}, we arrive at the Hamiltonian of the form
\begin{equation}
  \hat{H}_{2e}=\sum_{ij}\hat{d}^{\dagger}_{i}\braket{\psi_i|\hat{H}}{\psi_j} \hat{d}_{j} +\frac{1}{2}\sum_{ijkl}\hat{d}^{\dagger}_{i}\hat{d}^{\dagger}_{j}\hat{d}_{k}\hat{d}_{l}V_{ijkl},
  \label{Htwo}
\end{equation}
where $\hat{d}^{\dagger}_{i}$ creates an electron in the $i$-th spin-orbital. The one-electron energy is taken into account with the matrix elements $\braket{\psi_i|\hat{H}}{\psi_j}$. 
The Coulomb matrix element 
\begin{equation}
V_{ijkl}=\kappa\braket{\psi_{i}(\mathbf{x_1})\psi_{j}(\mathbf{x_2})\frac{1}{|\mathbf{r_{12}}|}}{\psi_{k}(\mathbf{x_1})\psi_{l}(\mathbf{x_2})}, 
\label{coulF}
\end{equation}
for the single-electron wave functions given by linear combinations of atomic orbitals $p_z$
give
\begin{eqnarray}
V_{ijkl}&=&\kappa\langle\psi_{i}({\bf x_{1}})\psi_{j}({\bf x_{2}})|\frac{1}{|{\mathbf{r_{12}}}|}|\psi_{k}({\bf x_{1}})\psi_{l}({\bf x_{2}})\rangle\nonumber \\
&=&\kappa \sum_{\substack{a,\sigma_{a};b,\sigma_{b};\\ c,\sigma_{c};d,\sigma_{d} }}\beta_{a,\sigma_{a}}^{i*}\beta_{b,\sigma_{b}}^{j*}\beta_{c,\sigma_{c}}^{k}\beta_{d,\sigma_{d}}^{l}\delta_{\sigma_{a};\sigma_{d}}\delta_{\sigma_{b};\sigma_{c}}\times \nonumber\\&&\langle p_{z}^{a}({\bf r}_{1})p_{z}^{b}({\bf r_{2}})|\frac{1}{|{\mathbf {r_{12}}}|}|p_{z}^{c}({\bf r_{1}})p_{z}^{d}({\bf r_{2}})\rangle.
\end{eqnarray}
For the Coulomb integral we apply the two-center approximation \cite{2c}
$ \langle p_{z}^{a}({\bf r}_{1})p_{z}^{b}({\bf r_{2}})|\frac{1}{|{\mathbf {r_{12}}}|}|p_{z}^{c}({\bf r_{1}})p_{z}^{d}({\bf r_{2}})\rangle=\frac{1}{r_{ab}} \delta_{ac}\delta_{bd}$ for $a\neq b$ and for the single center integral $a=b$
we take 16.522 eV \cite{tbci2}.

%\begin{equation}
% E_C=\kappa\braket{\Psi}{  |\mathbf{r_{12}}|^{-1} | \Psi }
%\end{equation}

% \begin{equation}
%  \kappa=\frac{e^2}{4\pi\epsilon\epsilon_0}
% \end{equation}

 \begin{figure}

  \includegraphics[scale=0.25]{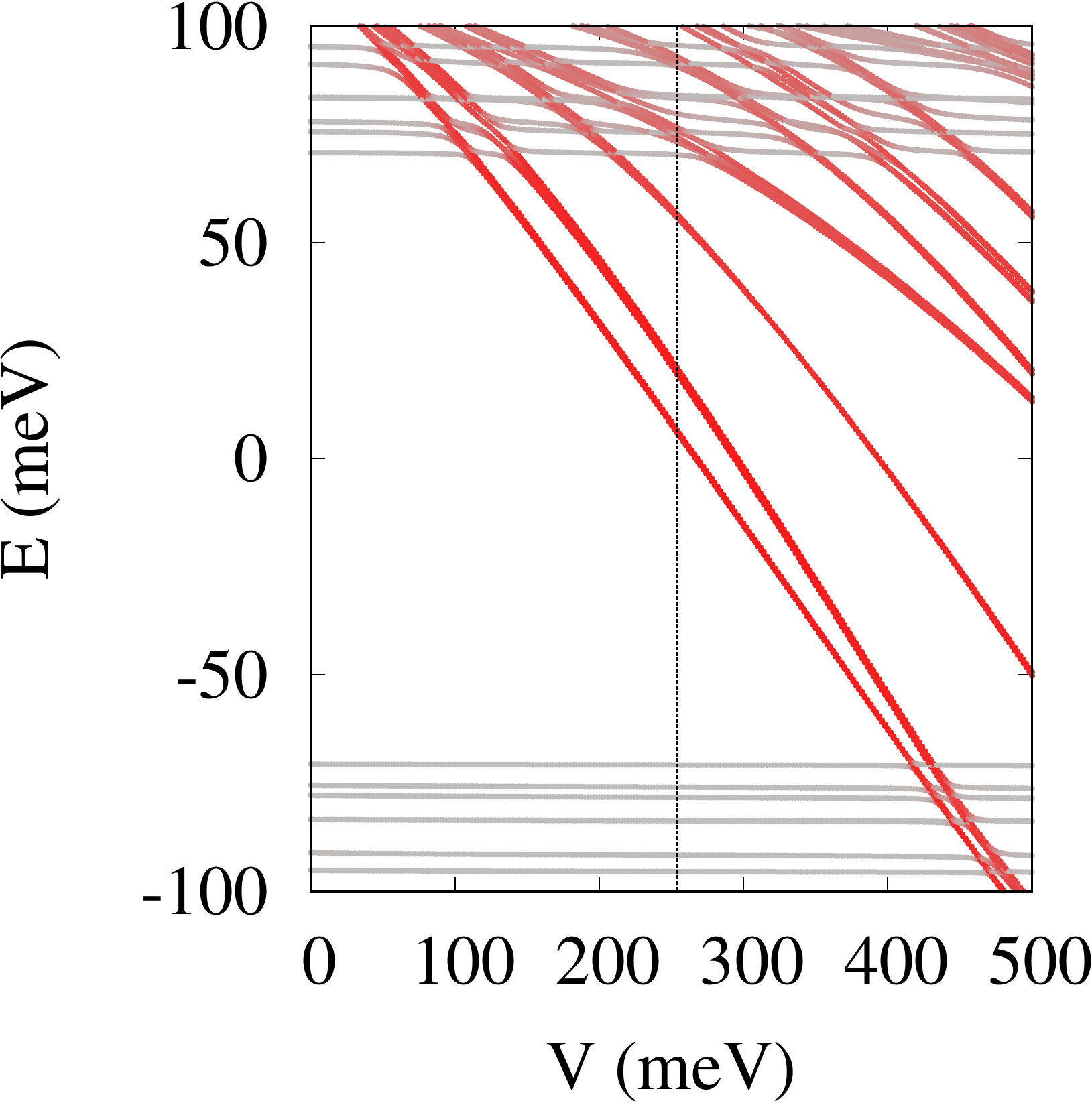} 
 
  \caption{The energy spectrum as a function of the quantum dots depth $V$ for the interdot distance of $2d=20$ nm at zero magnetic field.
  The red (grey) color of the curves indicates  the localization of electron density inside (outside) the quantum dots. 
  The dashed line shows the  potential depth chosen for the further calculations. }

  \label{stany11}
 \end{figure}

\begin{figure}[h!]
 
\begin{tabular}{c c}
(a)\hspace{-0.3cm} \includegraphics[scale=0.24]{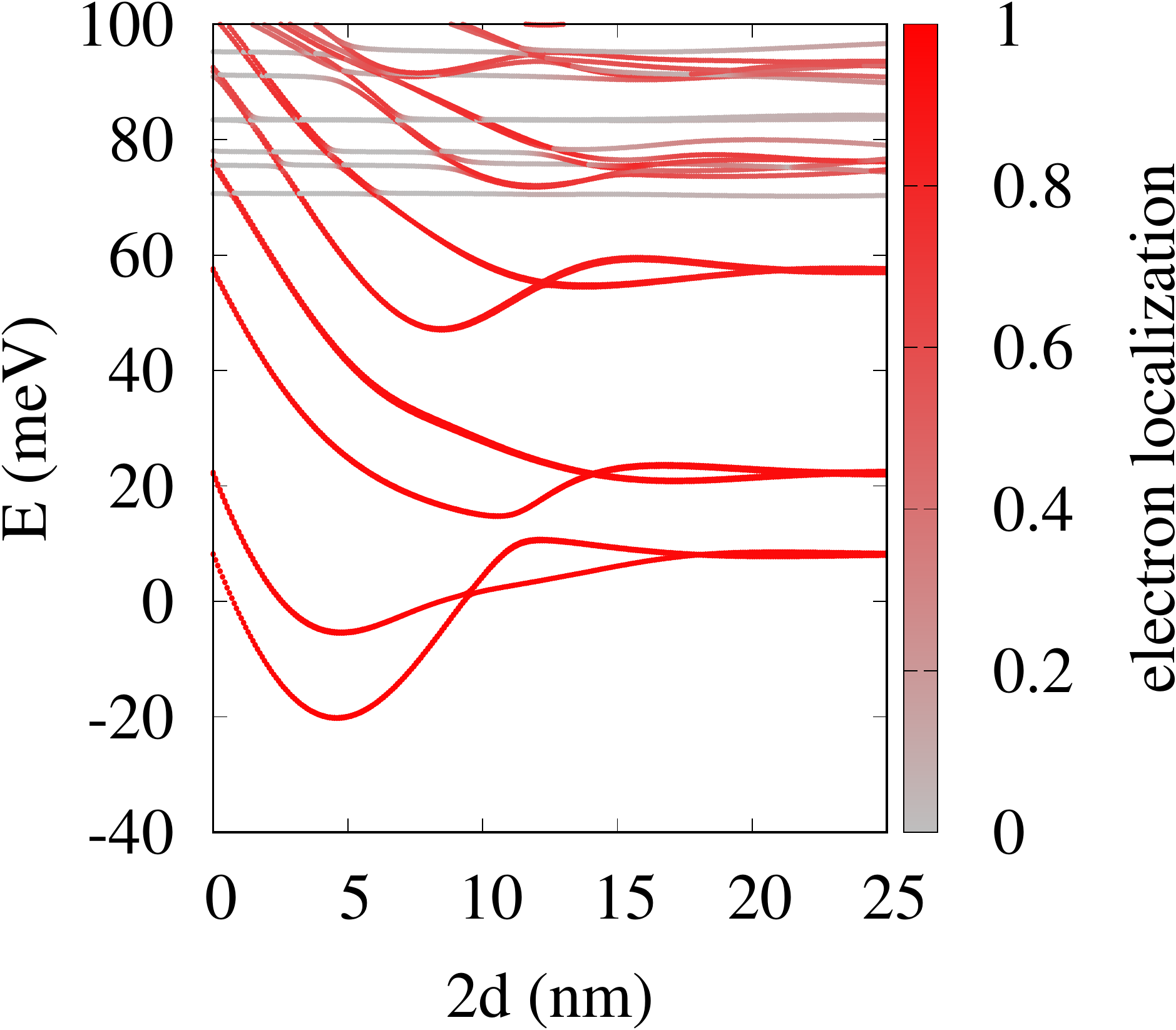} & (b)\hspace{-0.3cm}\includegraphics[scale=0.24]{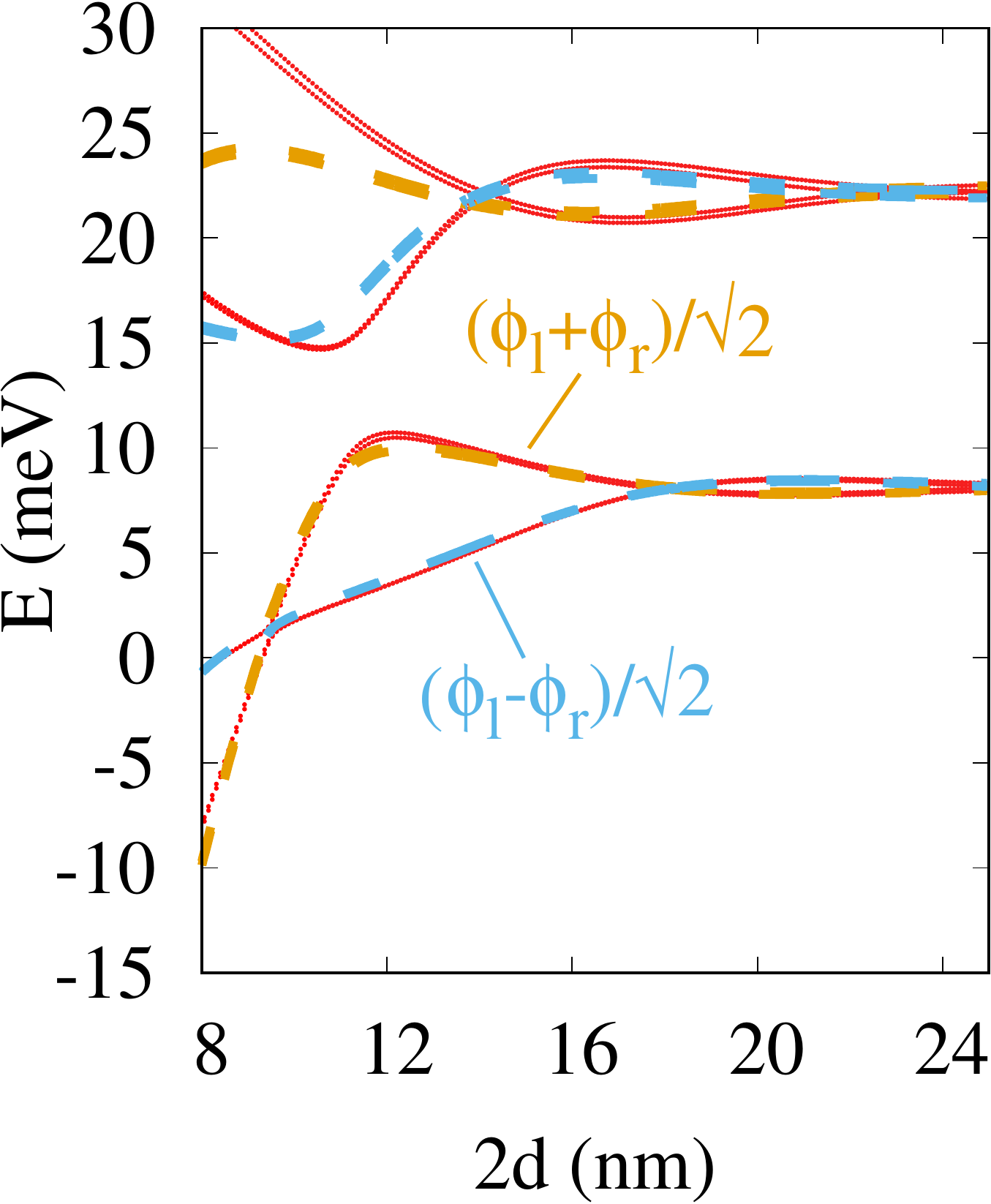} \\
(c)\hspace{-0.3cm} \includegraphics[scale=0.24]{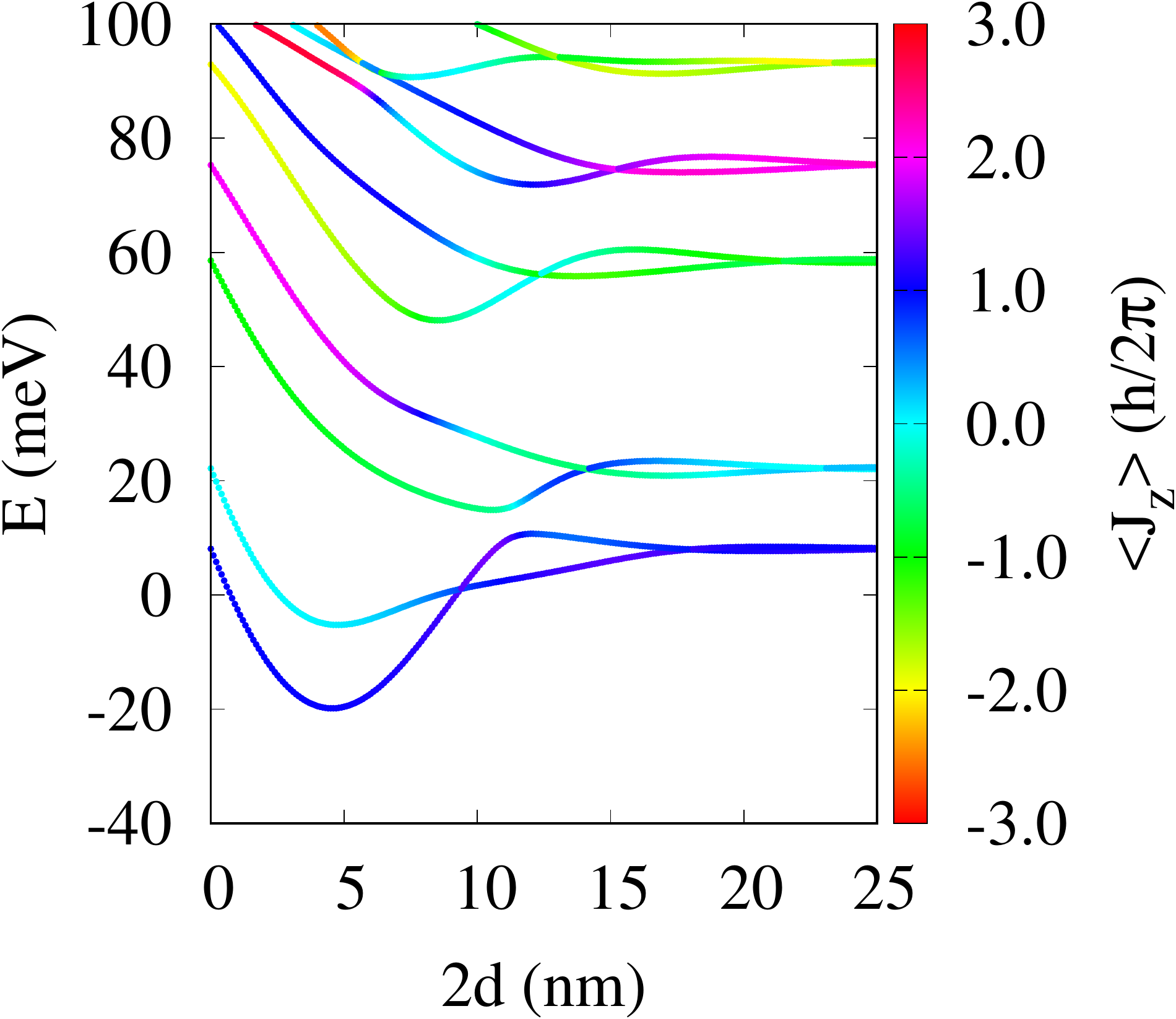}& (d) \hspace{-0.3cm}\includegraphics[scale=0.24]{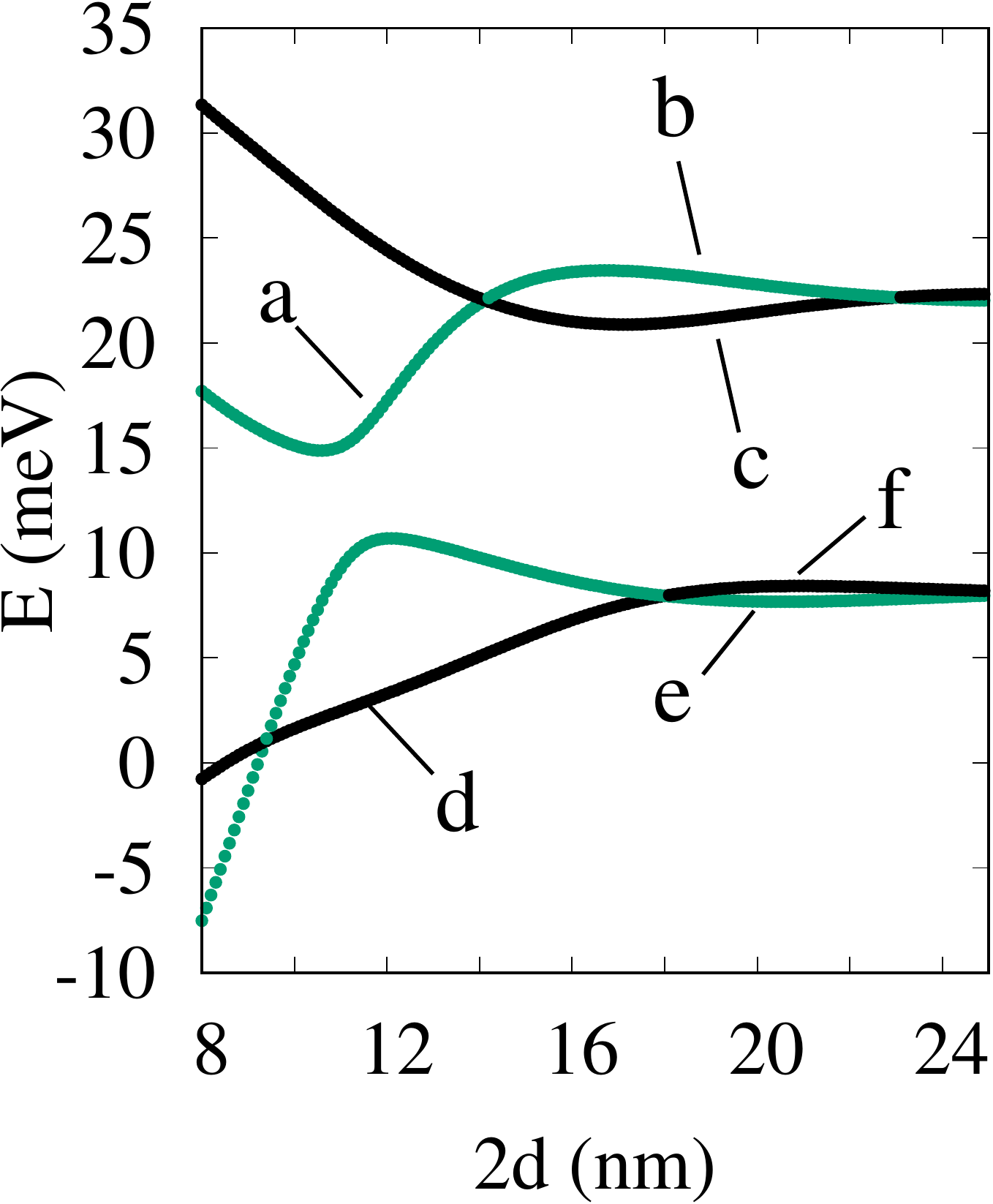}\\
\hspace{-1.2cm}(e)\hspace{-0.1cm} \includegraphics[scale=0.24]{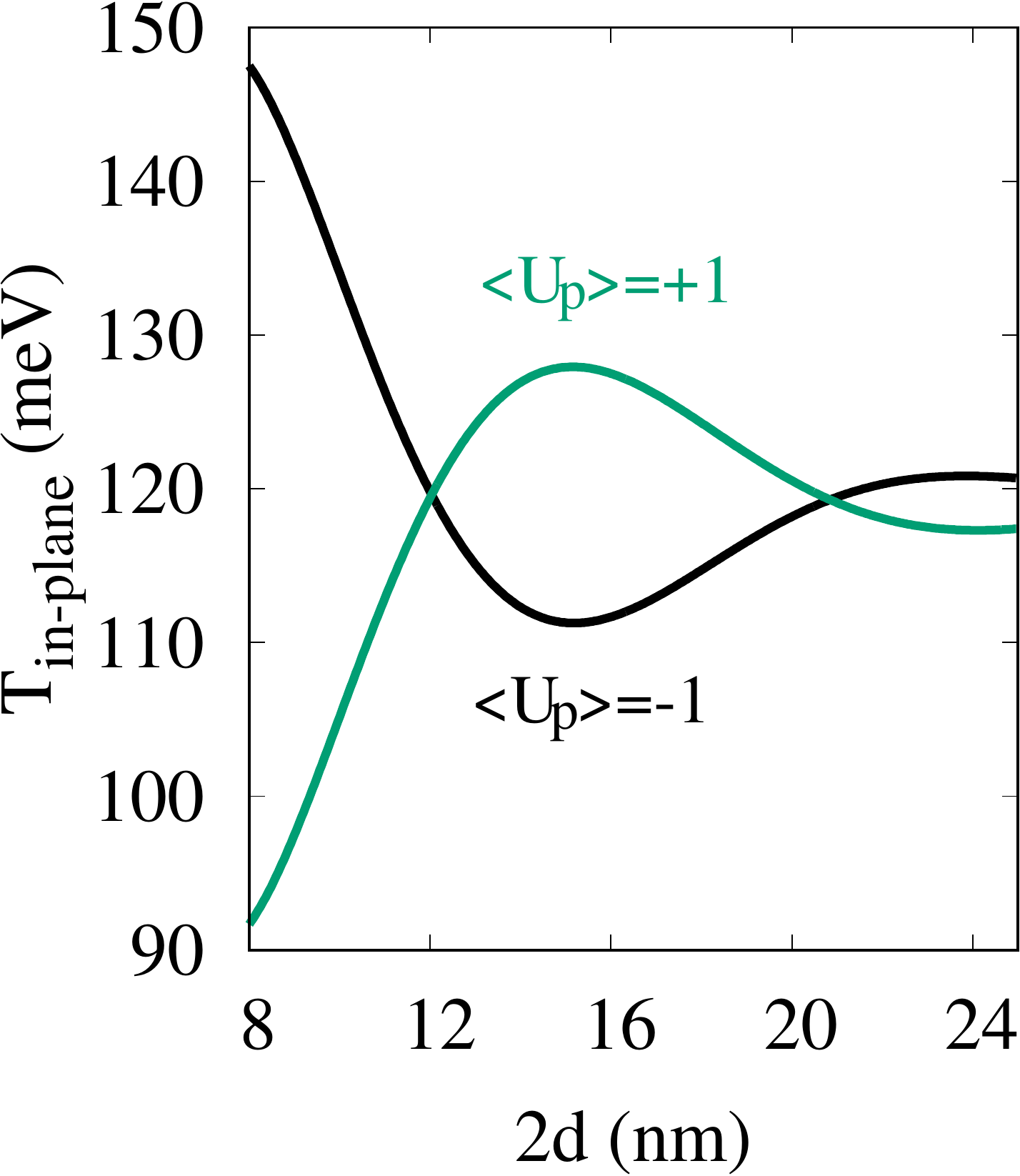}& (f) \hspace{-0.3cm}\includegraphics[scale=0.24]{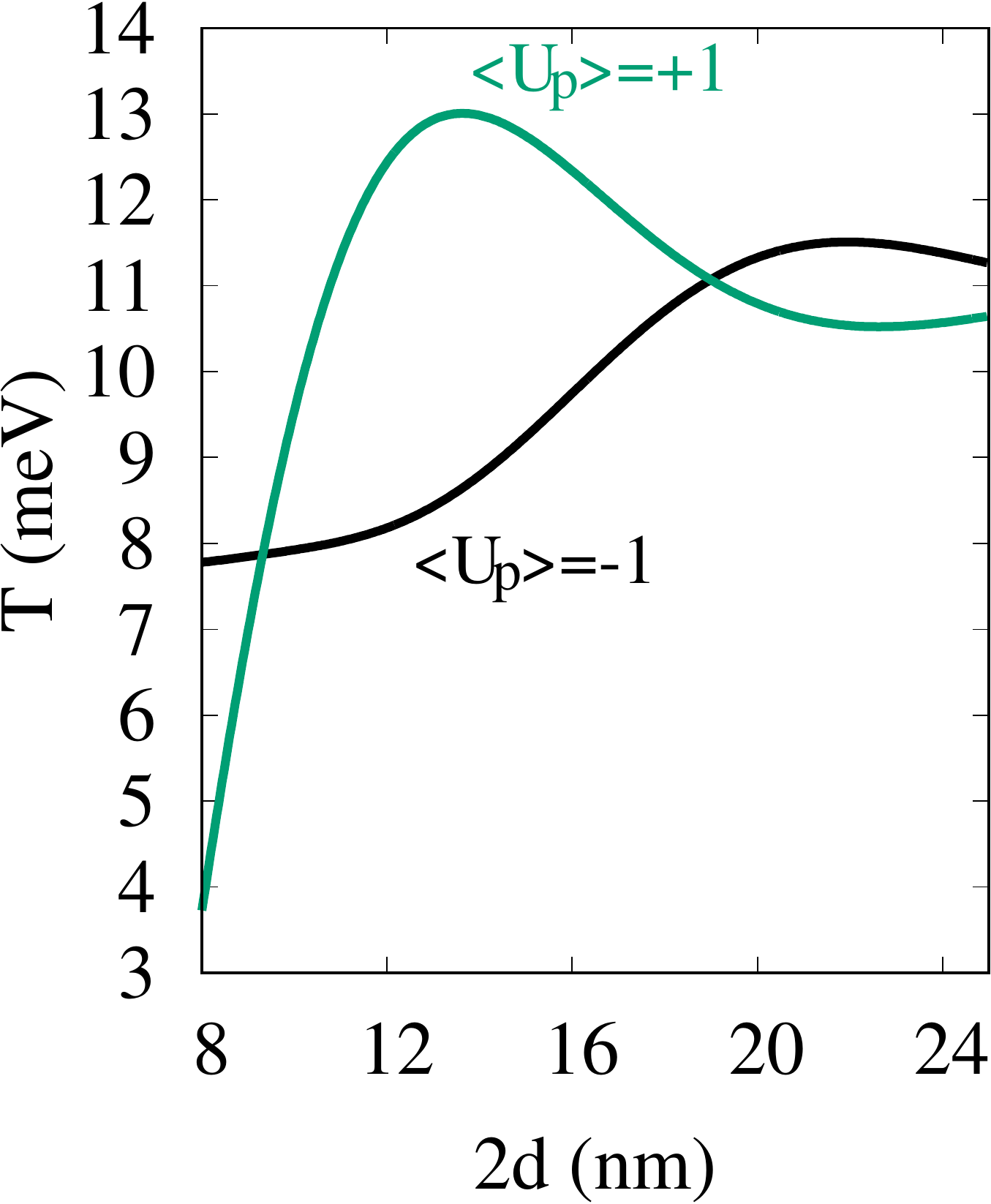}\\
\end{tabular}
 \caption{Single electron spectrum as a function of the interdot distance as obtained by the atomistic tight-binding (a,b) and by the low-energy 
continuum approximation (c,d).  (a) The color of the lines indicates the electron localization -- with the charge density integrated 
within the the distance of $2R$ from the centers of the dots. (b) Zoom of (a) -- the red lines are the same as in (a) while the orange and blue ones
have been obtained as a sum or difference of the wave functions localized in single dots (see text). 
(c) The energy levels obtained by the low-energy approximation with the color indicating the average value of the total angular momentum \reff{angmom}.
(d) Same as (c) only with the color of the line indicating the eigenvalue of the $\hat{U}_P$ operator: the green one +1, and the black one -1.
(e) The in-plane kinetic energy for the lowest-energy $\hat{U}_P$ eigenstates The average value of the off-diagonal part of Hamiltonian (6) with $t_\perp$ excluded.
(f) The kinetic energy for the  $\hat{U}_P$ eigenstates with both intralayer ($\pi$ and interlayer hopping ($t_\perp$). The average value of the entire off-diagonal part of Hamiltonian (6).
}
 \label{stany1}
\end{figure}

\section{Results}

\begin{figure*}[htbp]
\setlength{\tabcolsep}{10pt}
\begin{tabular}{c c c}
a)\includegraphics[scale=0.15]{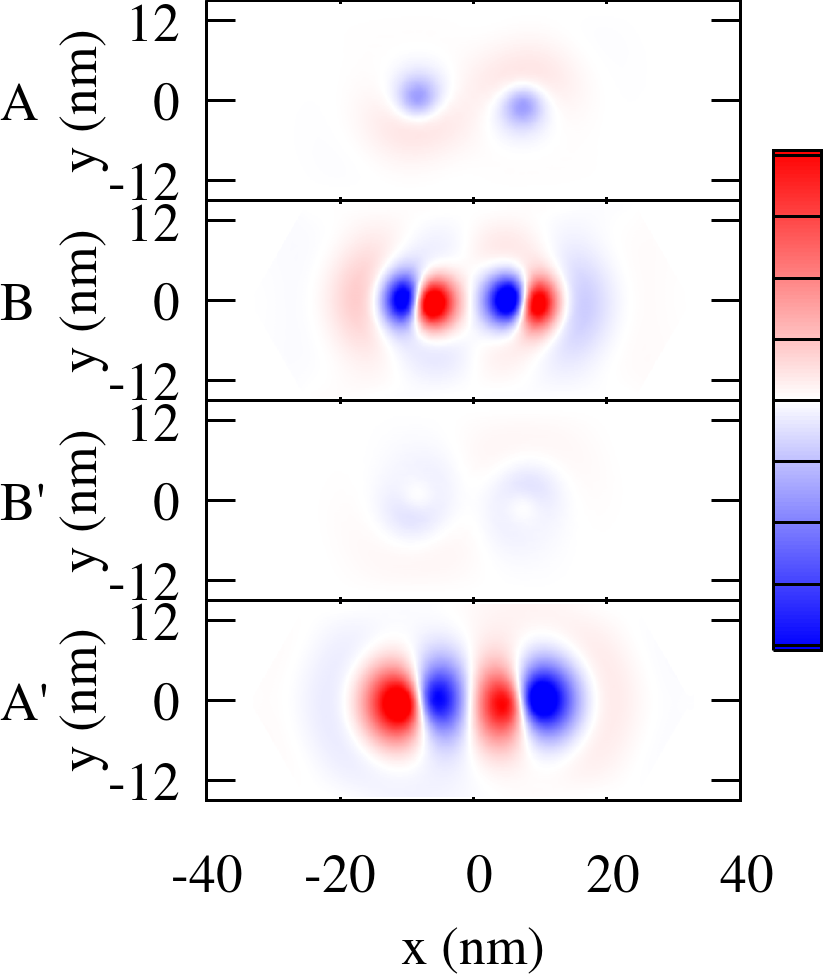} & b)\includegraphics[scale=0.15]{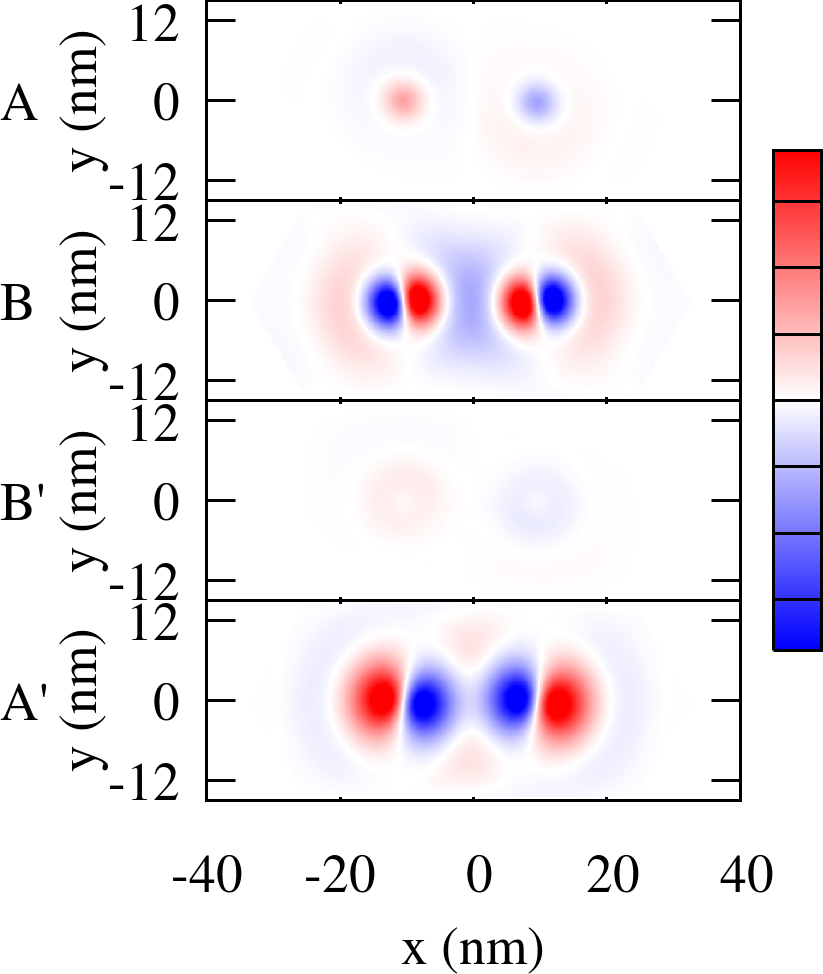} &c)\includegraphics[scale=0.15]{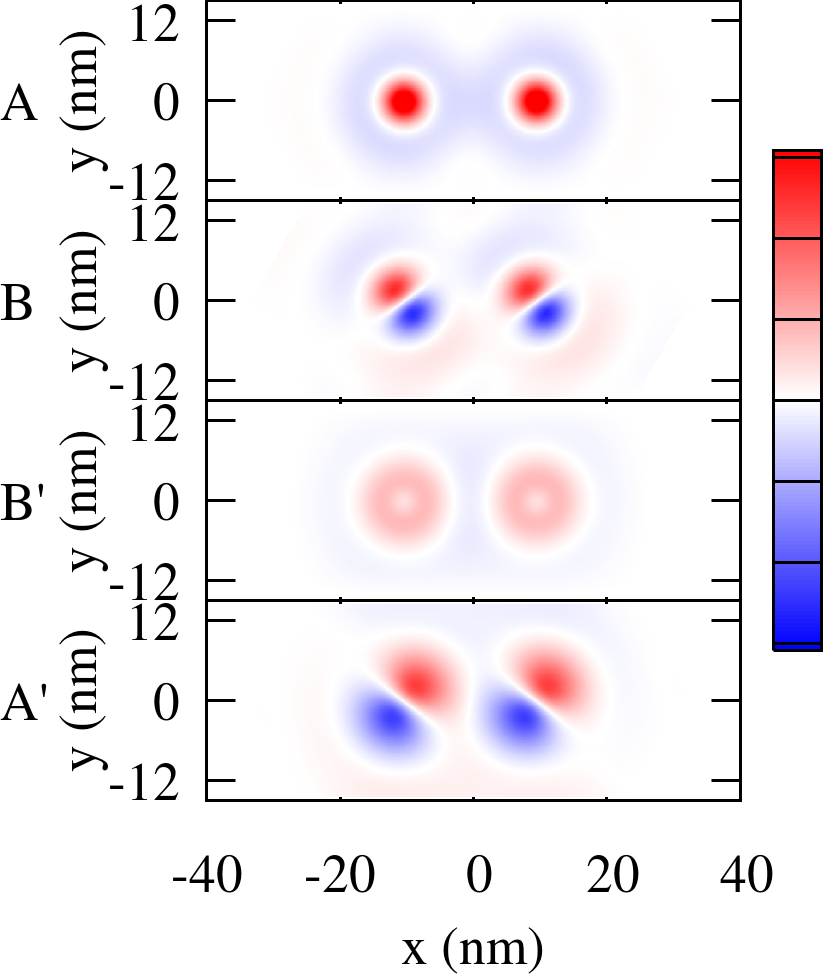}\\
% a)&b)&c)\\ 
d)\includegraphics[scale=0.15]{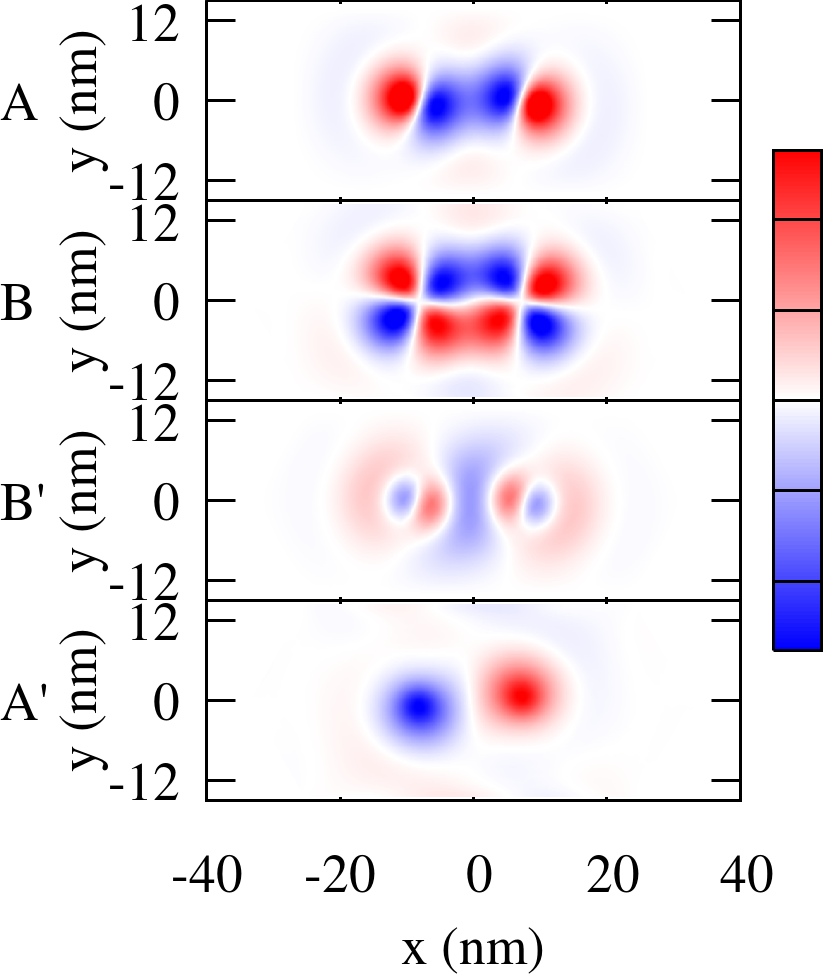} & e)\includegraphics[scale=0.15]{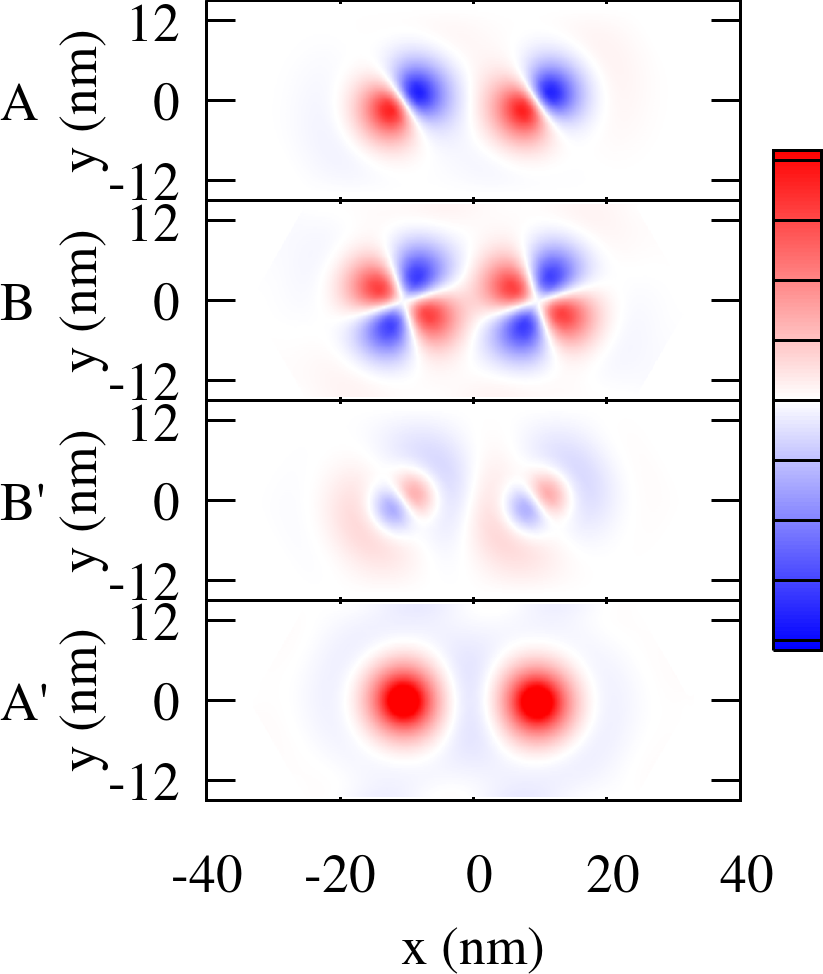}  &f)\includegraphics[scale=0.15]{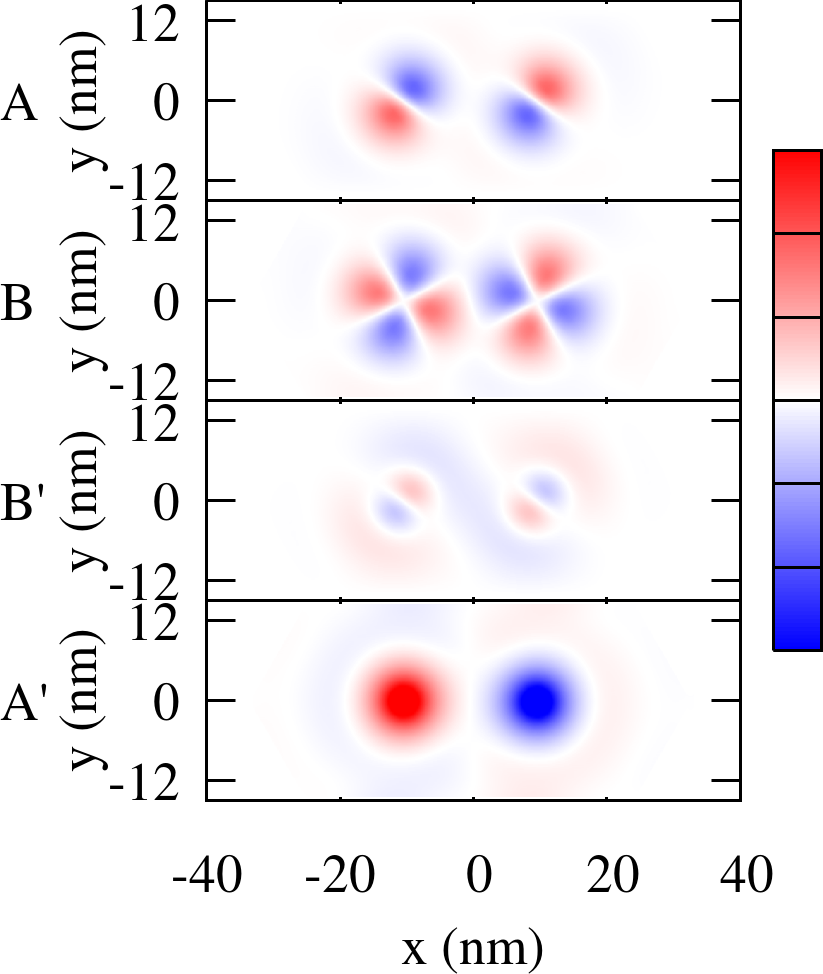}\\
% d)&e)&f) 
\end{tabular}
 \caption{Real part of the wave functions for the energy levels indicated on the Fig. \ref{stany1}(d) at the sublattices of both layers by the corresponding letters. 
 The states labeled as a) and d) correspond to $2d=8$ nm and the rest of the plots  to $2d=20$ nm.
The deepest red (blue) color indicate the most positive (negative) value of the real part of the wave function 
and the white color indicates its vanishing value.}
 \label{Wfal1}
\end{figure*}
 \begin{figure}
\begin{tabular}{c c}
(a)\hspace{-0.4cm} \includegraphics[scale=0.235]{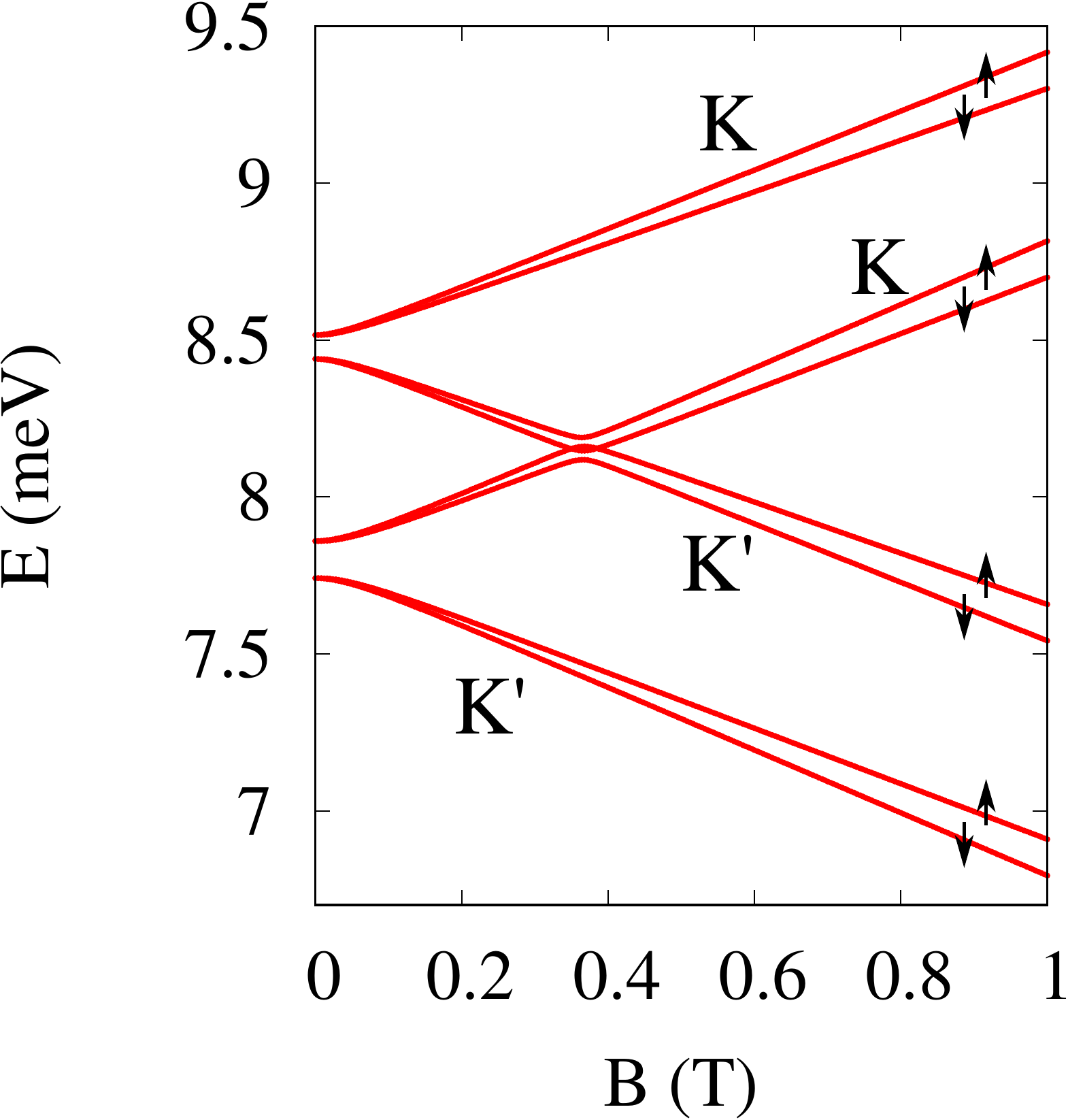} & (b) \hspace{-0.4cm} \includegraphics[scale=0.235]{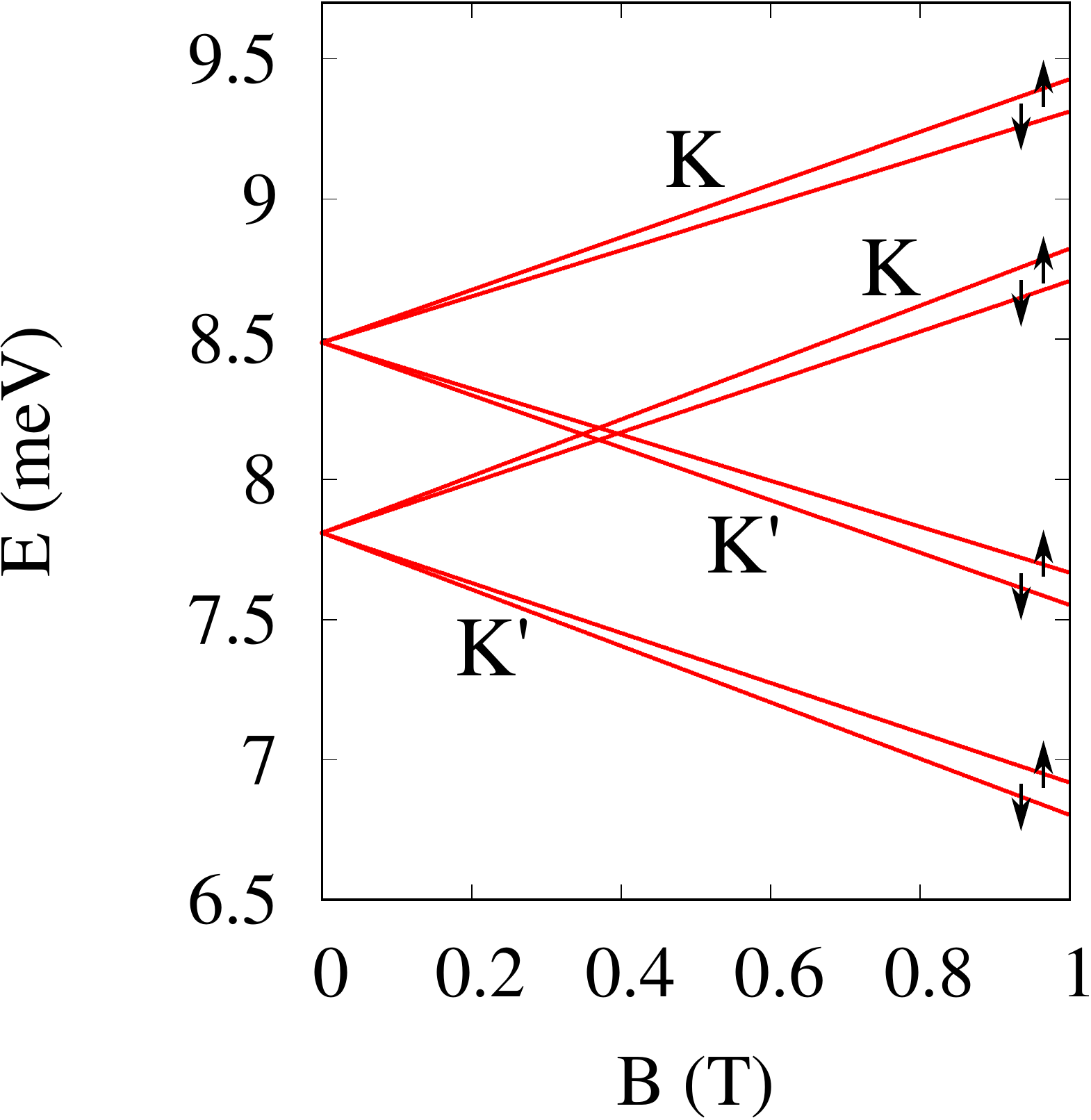}\\
 
\end{tabular}
   \caption{Eight lowest eigenvalues localized in the dot as a function of perpendicular magnetic field
for $2d=20$ nm as obtained by the tight-binding method. The (a) and (b) correspond to a smaller and larger flake respectively. The avoided crossings in (a) result from the valley mixing
by the armchair edge.}
  \label{inB}
 \end{figure}

 \subsection{Single-electron energy levels}

 The single-electron spectrum obtained with the atomistic tight-binding is displayed in Fig. \ref{stany11} as a function  of the depth of the dots for the interdot distance $2d=20$ nm for zero magnetic field. 
  The color of the lines indicates the extent of the electron localization within QDs -- the red
 ones are localized in the QD potentials, and the grey energy levels correspond to the states which are localized outside  the dots. Fig. \ref{stany11} shows that the latter ignore the QD potential.
 The vertical line in Fig. \ref{stany11} indicates the potential value $V=250$ meV that is taken for further calculations.  
 Each of the energy levels plotted in Fig. \ref{stany11} is nearly fourfold degenerate with respect to the spin and valley. 
 
For identical quantum dots the splitting between the two lowest energy levels is a result of the interdot coupling. 
 These  energy levels tend to a further degeneracy at large $d$ -- see Fig. \ref{stany1}.
 Typically, in the coupled quantum dots for small $d$ the bonding and antibonding electron orbitals
 are formed and the splitting  energy in III-V systems  is a monotonic function of $d$ \cite{burkard}. 
This is not the case that we observe in Fig. \ref{stany1}(a) 
 -- where crossings of the two lowest energy levels are seen. 

 The energy spectrum obtained with  the  continuum approximation of Fig. \ref{stany1}(c,d) is similar
 to the exact results of Fig. \ref{stany1}(a,b). 
 In Fig. \ref{stany1}(c)  we  plotted by the color
 of the line the average value of the total angular momentum (\ref{angmom}) calculated for the envelope function
 of the continuum Hamiltonian. At large $d$, when the interdot tunneling
 is negligible and for $d=0$, where the potential has a rotational symmetry, $\hat J_z$ eigenvalue is a good quantum number.
  The separate components of the envelope wave function 
 correspond then to the angular momentum quantum number \cite{biqd0} of $(m,m+1,m,m-1)$ for sublattices A,B,B' and A' respectively. 
 The ground state for the circular QD corresponds to $m=1$ and the lowest excited state with $m=0$ in accordance with Ref. \cite{biqd0}.

 For a general interdot distance $d$  the total angular momentum is no longer quantized.
However, each of the components has  a strict symmetry with respect to the point inversion through the center between the dots.
 The total wave function is an eigenstate of $\hat{U}_P$ and the eigenvalues $\pm 1$ are marked in Fig. \ref{stany1}(d) with the color of the line.

 In order to analyze in a more detail  formation of the extended orbitals by the single-dot wave functions 
 we constructed the localized (ionic) orbitals of  the left $l$ and right $r$ quantum dots using the tight-binding approach. 
 The wave functions $l$ and $r$ were constructed separately from 1) the two lowest energy states  $\phi_1$, $\phi_2$ and 2) the third and fourth states $\phi_3$, $\phi_4$. 
 The ionic functions were taken as a superposition of the Hamiltonian eigenstates,
  $l_{1,2}=(\phi_1+\exp(i\alpha)\phi_2)$,
 $r_{1,2}=(\phi_1+\exp(i(\alpha+\pi))\phi_2)$, where the phase $\alpha$ was taken to maximize the electron localization at the left or right side  of the origin.
 Then for Fig. \ref{stany1}(b) the extended states were produced as a constructive and destructive interference of the $l$ and $r$ functions
 $\phi^{1,2}_c=\frac{1}{\sqrt{2}} \left(l_{1,2}+r_{1,2}\right)$ and $\phi_d^{1,2}=\frac{1}{\sqrt{2}} \left(l_{1,2}-r_{1,2}\right)$, respectively.
 A similar operation was performed starting from the third and fourth energy level wave functions  $\phi_3$ and $\phi_4$.
 The mean values of the energy calculated for $\phi_c$ and $\phi_d$ wave functions were plotted by blue and orange lines in Fig. \ref{stany1}(b), respectively.
The red line shows the exact result that are the same as in Fig. \ref{stany1}(a).  Note that, there is a one to one correspondence between the $\hat{U}_P$ eigenvalue and the superposition sign
 taken in Fig. \ref{stany1}(b).

  For the scalar electron envelope function in a III-V material  $\phi_c$ and $\phi_d$ would correspond simply to the bonding and antibonding orbitals.
For multicomponent wave functions of opposite symmetry the situation
is more complex.
In particular, for holes in the artificial molecules formed by vertical quantum dots \cite{ab1,ab2,ab3,rev1} the wave functions need to be described by multicomponent wave function for each of the 
valence bands that become degenerate at the $\Gamma$ point.
In these systems the parity of the heavy hole component is opposite
to the light hole component. For large interdot distances 
the interdot tunneling is carried by the light hole component
a bonding state at the light hole component is formed which triggers
an antibonding heavy hole ground state  \cite{ab1,ab2,ab3,rev1}. 
In III-Vs' the crossing of the states of opposite parities is observed only
for the vertical \cite{ab1,ab2,ab3}  and not the lateral couping of the quantum dots \cite{rev1}.

For the lateral DQD studied here the origin of the non monotonic behaviour of the spacing between the lowest  energy levels of Figs. \ref{stany1}(c,d)
 can be understood by inspection of the wave function. In Fig. \ref{Wfal1} we plotted the real part of the wave functions for energy levels 
 and $d$ values marked by the corresponding letters in Fig. \ref{stany1}(d).
 For the plots at large interdot distances [cf. Fig. \ref{Wfal1}(c,e,f)]
 -- each of the wave function components 
 depends on the azimuthal angle as $\Re\left[\exp(iL\Phi)\right]$ as for
 the angular momentum eigenfunctions with the quantum number $L$.
 In particular, in the lowest-energy state that is twofold degenerate at large $d$ [cf. Fig. \ref{Wfal1}(e,f)] the total angular momentum quantum number is  $m=1$, and the $\hat{U}_P$ eigenvalue is +1 in Fig. \ref{Wfal1}(e) and -1 in Fig. \ref{Wfal1}(f).
 As the interdot distance is varied [cf. Fig. \ref{stany1}(d)] the energies of the states change
  but in a complex
 manner since two components of the wave functions are even in terms of  $\hat{U}_P$ operator while the other two are odd, and the two parities
 have opposite consequences for  energies  as $d$ is modified.
 Moreover,  
 the contributions of the components change. For instance: the state with the wave function given in  Fig. \ref{Wfal1}(f) has a dominant A' component which is antibonding.  
 For lower $d$ the A and B components  [Fig. \ref{Wfal1}(d)]  are increased at the expense of A' one.
The black energy level
 in Fig. \ref{stany1}(d) becomes the ground state level between the interdot distance 8 nm and 16 nm. 
 The varied contributions of the components involve avoided crossings between the energy levels of the same $\hat{U}_P$ parity
 -- and the identity of the lines that avoided cross can be traced by the average angular momentum [Fig. \ref{stany1}(c)].
 At $d=0$ the sequence of $J_z$ eigenvalues in the order of the energy is the same as in the large $d$ limit.

In order to analyze the effect of the molecular coupling on the energy levels
and the crossing observed in Fig. \ref{stany1}(a-d) we plotted in Fig. \ref{stany1}(e) the contribution of the in-plane kinetic energy 
-- for the hopping between A-B and A'-B' sublattices within each layer that are due to the $\pi$ operator in the Hamiltonian given by Eq. (6). 
Figure \ref{Wfal1}(d,e,f) indicates that for the ground state the coupling 
of the single-dot energy levels -- corresponding to $m=1$ each, 
appears for the largest interdot distances at the $B$ component which
corresponds to orbital angular momentum quantum number $L=m+1=2$. 
The state that forms a bonding component there [Fig. \ref{Wfal1}(e)] with a positive
eigenvalue of  $\hat{U}_P$ operator decreases in the energy [Fig. \ref{stany1}(e)] due to the interdot coupling. When the dots get closer 
the 
components with orbital angular momentum $L=m=1$ at A and B' sublattices form a molecular orbital which is antibonding (bonding) for the $\hat{U}_P$ eigenstate with eigenvalue +1 (-1). Due to the inverse contribution to the energy the first crossing at the ground-state energy level is observed near $2d=18$ nm [Fig. \ref{stany1}(a,d)]. 

The in-plane kinetic energies for the two states cross for even larger distance [Fig. \ref{stany1}(e)] of $2d=21$ nm. For the interlayer hopping  between A and B' sublattices included the crossing 
of kinetic energies 
  occurs closer  [Fig. \ref{stany1}(f)] to the intersection  
of the energy levels [Fig. \ref{stany1}(d)].

 The second crossing of the ground-state energy levels near
$2d=8$ nm results from the activation of the interdot tunnel coupling for the A' component of the angular momentum $L=m-1=0$ [Fig. \ref{Wfal1}(d-f)]. 
Concluding, the ground-state energy level crossings for the states of opposite $\hat{U}_P$  parity results from the subsequent switching of the interdot
tunnel coupling for separate  components of the wave function at the sublattices.

 The magnetic field dependence of the energy spectrum is given in Fig. \ref{inB}. At this energy scale
the results for the smaller [Fig. \ref{inB}(a)] and the larger flake [Fig. \ref{inB}(b)] become distinguishable. For the smaller flake
we notice avoided crossings between the energy levels of opposite valleys which result from the valley
mixing effect of the armchair edge of the flake [at $B\simeq 0$ and $B\simeq 0.4$ T]
 The valley mixing opens avoided crossing between the states of the same spin and  $\hat U^P$ eigenvalue,
 and outside of the avoided crossings the energy levels in Fig. \ref{inB}(a) can be attributed with the valley and spin quantum numbers.
For the larger flake the valley mixing effects are not resolved on this scale.

 \begin{figure}[htbp]
\begin{tabular}{c} 
  a)\hspace{-0.3cm} \includegraphics[scale=0.4]{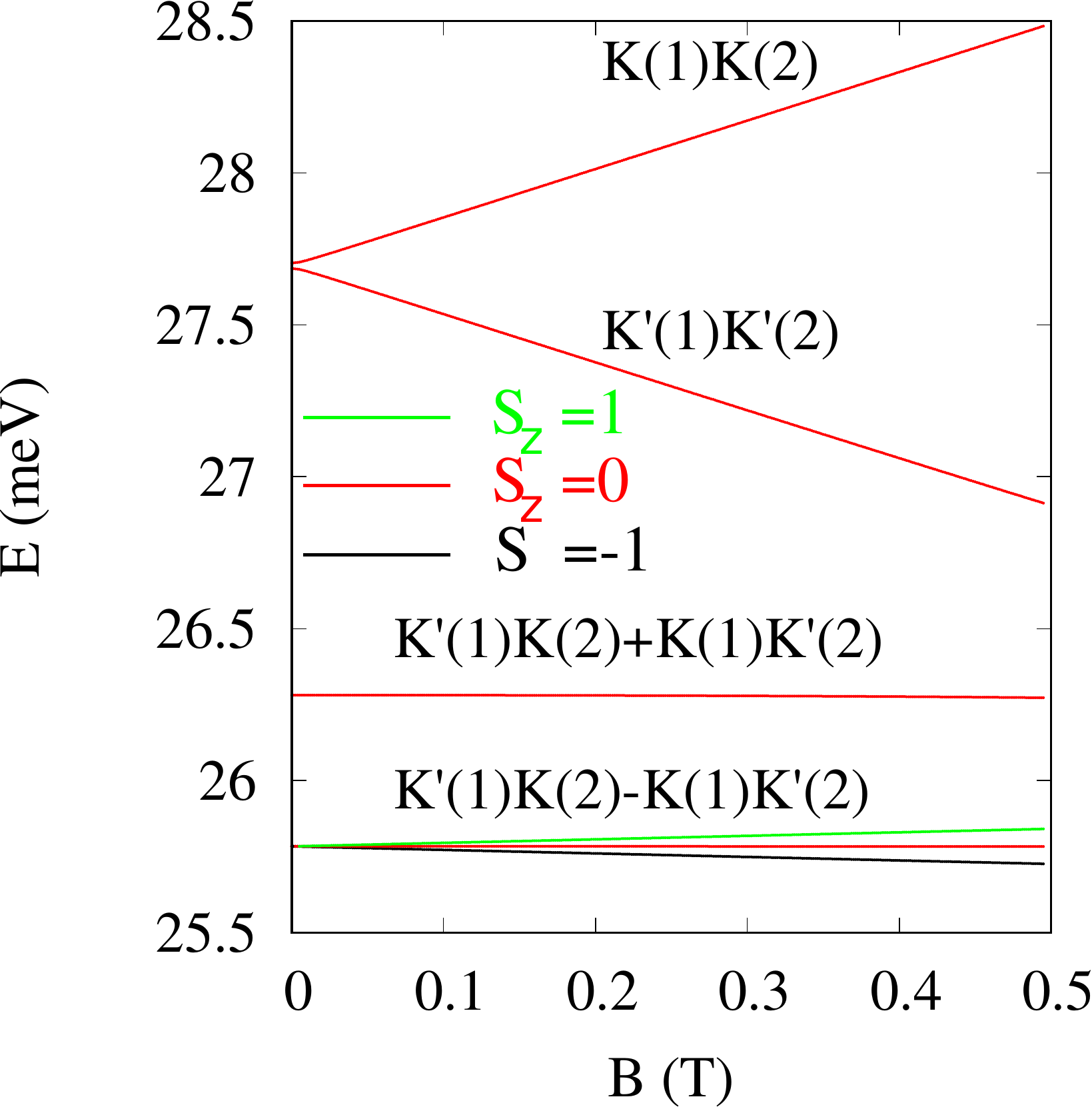} \\ b) \hspace{-0.3cm}\includegraphics[scale=0.4]{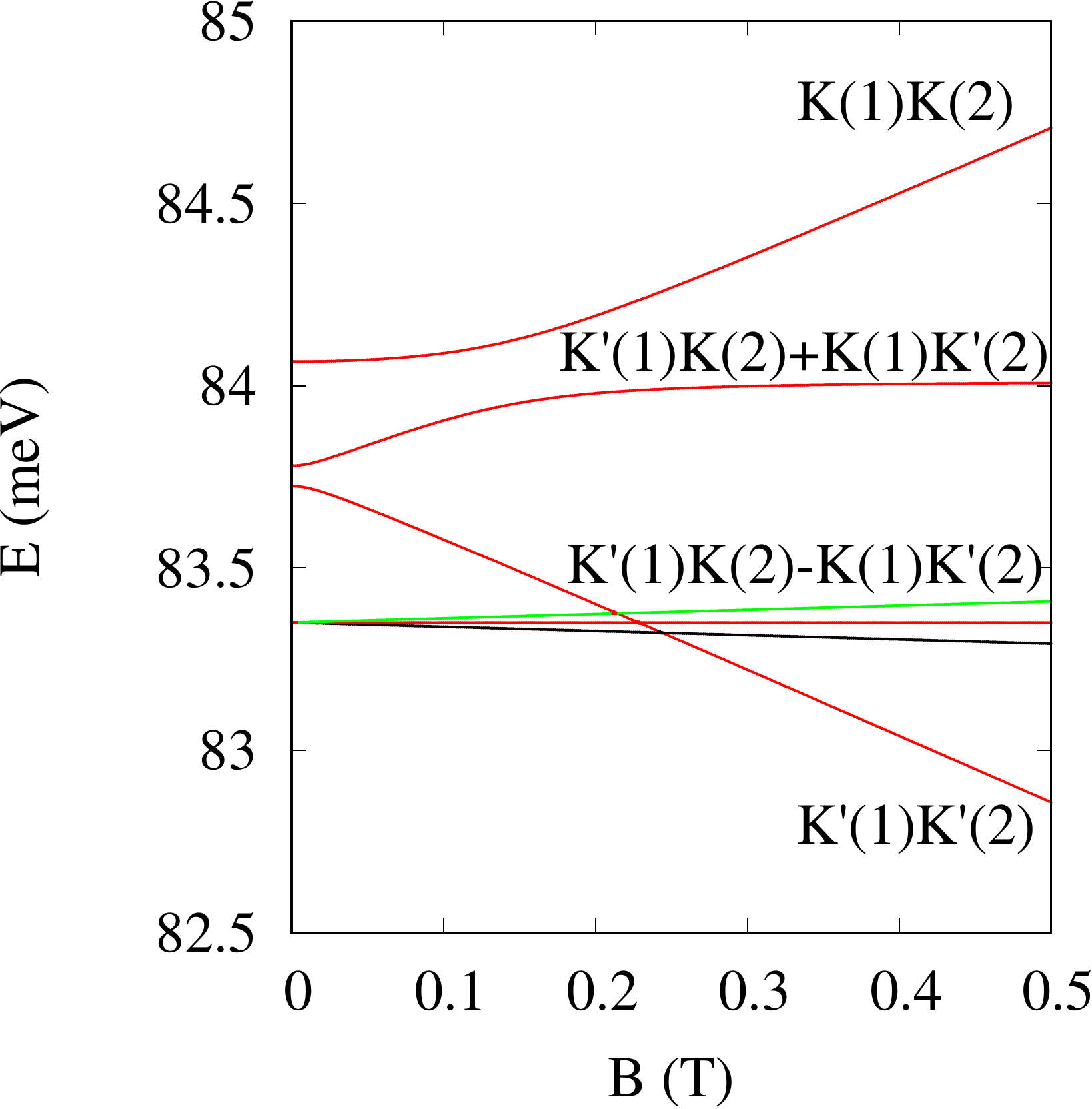}  
\end{tabular}
  \caption{(a) The low-energy spectrum  of the DQD potential for $2d=6$nm (b) Same as (a) but for 
a single QD defined within the flake. Both results obtained for the smaller flake.} \label{d6}
 \end{figure}

\begin{figure*}[htbp]
\centering
\includegraphics[scale=0.32]{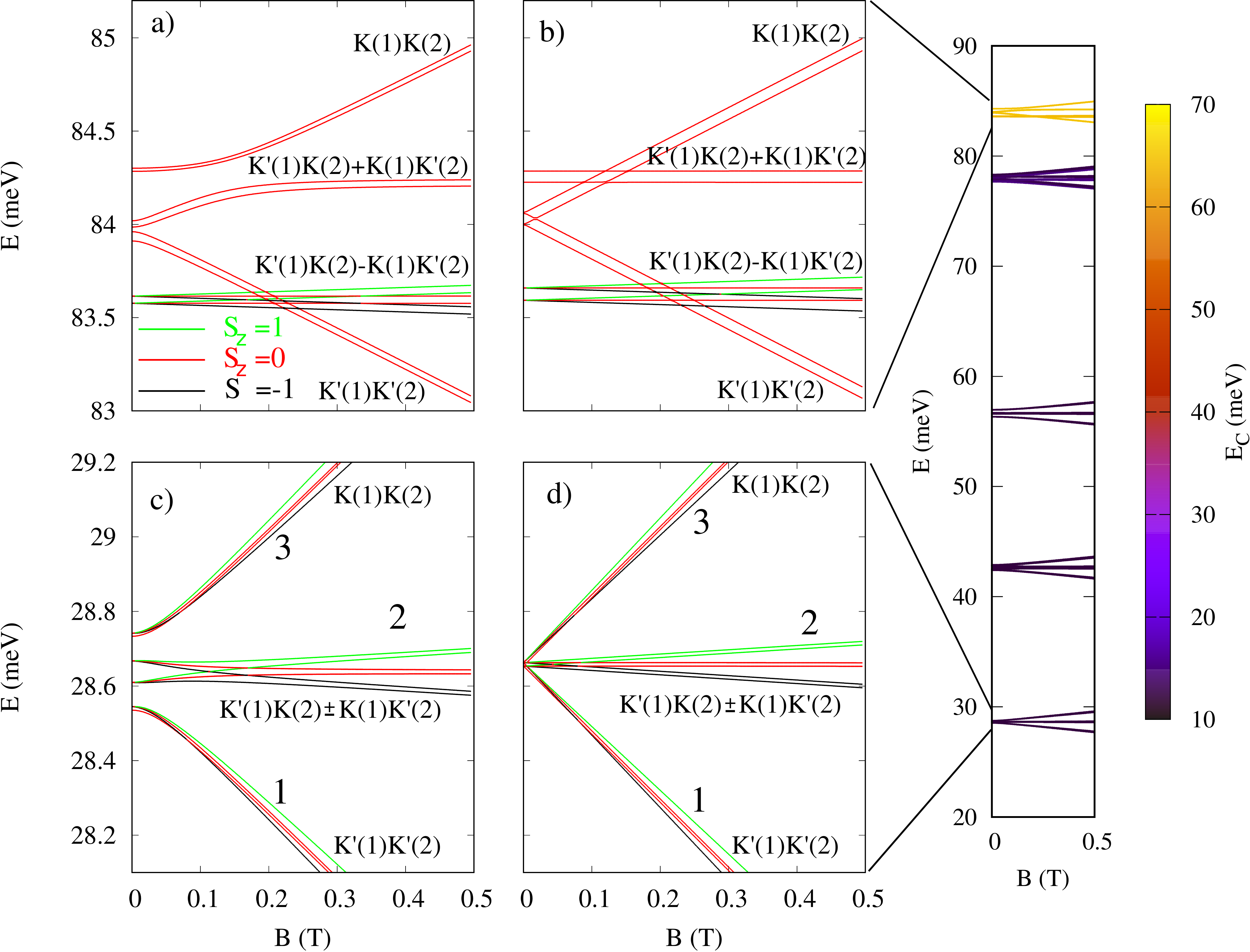}
 \caption{The two-electron energy spectrum for $2d=20$ nm. The right panel shows the groups of energy levels with the color
of the line indicating the expectation value of the electron-electron interaction energy. (a,b) - zoom on the high energy spectrum with electrons
in the same dot, (c,d) - zoom on the ground-state. In (a,c) the smaller flake, while in (b,d) the larger one was taken for calculations.}  \label{d20}
\end{figure*}

\begin{table*}
\begin{tabular}{l|c|c|c}
no. & $\Psi_s$ & $\Psi_v$ &$\Psi_\sigma$    \\ \hline
\hline 

\color{ForestGreen}16   & \color{ForestGreen}$l(1) r(2) -r(1)l(2)$ &\color{ForestGreen}$K(1) K (2)$ &$\uparrow(1)\uparrow(2)$\\
\color{red} 15    & \color{red}$l(1) r(2) -r(1)l(2)$ &$K(1) K (2)$\color{red} &$\uparrow(1)\downarrow(2)+\downarrow(1)\uparrow(2)$\\
 \color{red}14   & \color{red}$l(1) r(2) +r(1)l(2)$ &$K(1) K (2)$\color{red} &$\downarrow(1)\uparrow(2)-\uparrow(1)\downarrow(2)$\\
 13    & $l(1) r(2) -r(1)l(2)$ &$K(1) K (2)$ &$\downarrow(1)\downarrow(2)$\\

\hline
\color{ForestGreen}12     & \color{ForestGreen}$l(1) r(2) -r(1)l(2)$ &\color{ForestGreen}$K(1) K' (2)+K'(1) K (2)$ &\color{ForestGreen}$\uparrow(1)\uparrow(2)$\\
\color{ForestGreen}11     & \color{ForestGreen}$l(1) r(2) +r(1)l(2)$ &\color{ForestGreen}$K(1) K' (2)-K'(1) K (2)$ &\color{ForestGreen}$\uparrow(1)\uparrow(2)$\\

\color{red} 9-10    & \color{red}$l(1) r(2) -r(1)l(2)$ &\color{red}$K(1) K' (2)+K'(1) K (2)$ &\color{red}$\downarrow(1)\uparrow(2)+\uparrow(1)\downarrow(2)$\\
\color{red} 9-10     &\color{red} $l(1) r(2) -r(1)l(2)$ &\color{red}$K(1) K' (2)-K'(1) K (2)$ &\color{red}$\downarrow(1)\uparrow(2)-\uparrow(1)\downarrow(2)$\\

 \color{red}7-8    & \color{red}$l(1) r(2) +r(1)l(2)$ &\color{red}$K(1) K' (2)-K'(1) K (2)$ &\color{red}$\downarrow(1)\uparrow(2)+\uparrow(1)\downarrow(2)$\\
\color{red} 7-8    & \color{red}$l(1) r(2) +r(1)l(2)$ &\color{red}$K(1) K' (2)+K'(1) K (2)$ &\color{red}$\downarrow(1)\uparrow(2)-\uparrow(1)\downarrow(2)$\\
6     & $l(1) r(2) -r(1)l(2)$ &$K(1) K' (2)+K'(1) K (2)$ &$\downarrow(1)\downarrow(2)$\\
5     & $l(1) r(2) +r(1)l(2)$ &$K(1) K' (2)-K'(1) K (2)$ &$\downarrow(1)\downarrow(2)$\\

\hline 
\color{ForestGreen} 4   & \color{ForestGreen}$l(1) r(2) -r(1)l(2)$ &\color{ForestGreen}$K'(1) K' (2)$ &\color{ForestGreen}$\uparrow(1)\uparrow(2)$\\
\color{red} 3    & \color{red}$l(1) r(2) -r(1)l(2)$ &\color{red}$K'(1) K' (2)$ &\color{red}$\uparrow(1)\downarrow(2)+\downarrow(1)\uparrow(2)$\\
 \color{red} 2   & \color{red}$l(1) r(2) +r(1)l(2)$ &$K'(1) K' (2)$ &\color{red}$\downarrow(1)\uparrow(2)-\uparrow(1)\downarrow(2)$\\
 1    & $l(1) r(2) -r(1)l(2)$ &$K'(1) K' (2)$ &$\downarrow(1)\downarrow(2)$\\
\end{tabular}
\caption{The symmetries of the two-electron states, including the spatial $\Psi_s$, the valley $\Psi_v$ and the spin $\Psi_\sigma$ factors
where the total wave function $\Psi(1,2)=\Psi_s\Psi_v\Psi_\sigma$. The table corresponds to Fig. \ref{d20}(d), where the first column
orders the state in a growing energy order. Levels 7-8 and 9-10 are degenerate. The color of the lines corresponds to the $z$ component of the spin. The list
corresponds to the dominant contribution to the CI wave functions. \label{tablee}}
\end{table*}

\subsection{Two-electron spectra}
\subsubsection{Low $d$ limit}

The energy spectrum taken for a small interdot distance of $2d=6$ nm is displayed in Fig. \ref{d6}(a). 
At $B=0$ the ground-state is threefold degenerate.  One of the electrons occupies the $K$  and the other the $K'$ valley.  The valley degree of freedom  allows the electrons 
to acquire the same spatial distribution with the three possible components of the total spin. 
From the dominant contributions to the wave function one concludes that the approximate form of the wave function
for the three-fold degenerate ground state can be put in a separable form (normalization skipped) $\Psi_{gs}=\psi(1)\psi(2) \left(K(1)K'(2)-K'(1)K(2)\right) \Psi_T$,
where $\psi$ is the spatial orbital for the positive $\hat U_P$ eigenvalue, and 
$\Psi_T$ is one of the spin triplet wave functions for $S_z=-\hbar$, 0 or $\hbar$.
The triplet ground state at $B\neq 0$ is split only
by the Zeeman effect. The first excited state is non degenerate spin singlet that corresponds to electrons in opposite valleys
$\Psi_4=\psi(1)\psi(2) \left(K(1)K'(2)+K'(1)K(2)\right)\Psi_S$, where $\Psi_S=\left(\uparrow_1\downarrow_2-\uparrow_2\downarrow_1\right)$.
The second excited state at $B=0$ is two-fold degenerate. Both the electrons occupy either the $K'$ valley (the energy level that goes down in $B$),
$\Psi_5=\psi(1)\psi(2) K'(1)K'(2) \Psi_S$,
or the $K$ valley (the energy that goes up). Still the spatial orbital is the same for both electrons which is only possible due to the opposite
spins of the states -- they are both spin singlets. 
 
For the strong coupling case of Fig. \ref{d6}(a)
the two dots effectively form a single system and the spectrum resembles the one calculated for a single dot in Fig. \ref{d6}(b).
The difference for the circular dot is that the valley unpolarized singlet goes higher in the energy at $B=0$. The avoided
crossing with the $KK$ energy level as well as the avoided crossing between the two lower singlets at $B=0$ is due to the 
valley mixing effects of the armchair boundary condition. The ground state is still the triplet -- which was previously found
for a single-dot study of Ref. \cite{zarenia2e}.

\subsubsection{Weak coupling}

For electrostatic quantum dots the coupling between the dots is usually weak, so this case deserves a closer inspection
as the one which is the most likely to be encountered in an experimental situation. 
For a weak coupling of $2d=20$ nm the states can be separated into subgroups with both electrons in separate
dots or with both electrons in the same dot. The latter states appear higher in the energy spectrum and they are characterized
by a stronger electron-electron interaction. The right panel of Fig. \ref{d20} displays the energy spectrum in a wider energy range,
as a function of the external field, and the color of the lines display the average electron-electron interaction energy.
 In Fig. \ref{d20}(a,b) we display a zoom of the spectrum at the energy levels  with the electrons in the same dots,
for the smaller flake Fig. \ref{d20}(a) and for the larger one Fig. \ref{d20}(b).
The spectrum of Fig. \ref{d20}(a) is very close to the one displayed in Fig. \ref{d6}(b) -- for a single dot in a smaller flake. 
In Fig. \ref{d20}(b) the spectrum is similar, only the avoided crossing for opposite valleys are closed.
The closing of the avoided crossing in the two-electron spectra Fig. \ref{d20}(b,d) (for the larger flake)
with respect to Fig. \ref{d20}(a,c) (for the small flake) is of the same origin as closing the avoided crossing of the single-electron spectra of Fig. \ref{inB}(b) with respect to Fig. \ref{inB}(a), i.e. it results from removed coupling of the quantum-dot confined energy levels with the edge. The considered armchair edge of the flake is not equivalent with respect to the A and B, and respectively  A’ and B’ sublattices. The result of this non-equivalence is intervalley mixing of the states coupled to the edge. When the coupling is removed for a larger distance between the dots and the edge, the valley mixing disappears and the state of opposite valleys change their order in a crosssing instead of an avoided crossing.

The difference between the single QD [Fig. \ref{d6}(b)] result and the DQD [Fig. \ref{d20}(a)] is that for the latter,  the energy levels shift in pairs.
 The electron couple can be stored by the left $l$ or the right $r$ dot,
with the factor of the spatial wave function that can be put in one of the forms  $\left\{l(1)l(2)\pm r(1)r(2)\right\}/\sqrt{2}$.
The tunnel coupling between the dots leads to splitting of the energy levels into bonding and antibonding pairs 
that are observed in Fig. \ref{d20}(a).

The ground-state of the two-electron spectrum of Fig. \ref{d20} corresponds to separated electrons and is displayed in Fig. \ref{d20}(c) and (d)
for the smaller and for the larger flake respectively. 
Here, the two electrons occupy different spatial orbitals $l$ or $r$. For that reason the Pauli exclusion does not 
forbid them to occupy any of the spin-valley combinations, hence the ground-state at $B=0$ is nearly 16 times degenerate -- for the larger flake [Fig. \ref{d20}(d)].
For the smaller flake [Fig. \ref{d20}(c)] at $B=0$ the energy levels split into four quadruples. 
%The splitting results from the valley mixing avoided crossing that is present in Fig. \ref{inB}. The states at $B=0$
%are grouped into four nearly fourfold degenerate energy levels.
In the lowest (highest) quadruple the both electrons
levels occupy the lower (higher) energy level: they have the four spin states to occupy hence the number of the nearly degenerate states.
In both the lowest and highest quadruple a spin-singlet has a slightly lower energy than the spin triplet. 

For the larger flake as well as for the smaller one at higher $B$ the single-electron energy levels shift away from the valley mixing avoided crossing and the two-electron spectrum forms groups depending on the valley configurations. By analysis of the CI components of the wave function we found 
that the wave functions at high $B$ are approximately separable into product of spatial, valley and spin factors $\Psi(1,2)=\Psi_s\Psi_v\Psi_\sigma$
which are listed in Table \ref{tablee} for the energy order that is found in Fig. \ref{d20}(b) for $B=0.3$T.
In the group marked by 1 (3) both electrons occupy the $K'$ ($K$) valley.
Each group is formed by 4 energy levels with the splitting that is due to the Zeeman interaction. Since the valley degree of freedom
is frozen, the structure of each of the group is identical with the one found for two-electron quantum dots in III-V materials,
with the spin-singlet $S$ and spin-triplet energy level of zero $z$ component of the spin $T_0$ split by the exchange energy.
Here the splitting due to  exchange energy is of the order of 9 $\mu$eV. 

In the central group of energy levels -- marked by $2$ in Fig. \ref{d20}(b) the electrons occupy opposite valleys - hence
a nearly constant dependence on $B$, which here is only due to the Zeeman interaction. 
Each of the $S_z=0$ states is two-fold degenerate (energy levels 7-10 in Table \ref{tablee}) -- only the symmetry of the spatial wave function with respect to the 
interchange of electrons influences the energy of the state via interdot tunneling effect, and there are two
 spin and valley factors for both symmetric and antisymmetric states.
For each pair of energy levels of group (2) at high $B$ the one that is lower in the energy corresponds to a symmetric spatial wave function. 
The splitting energy is nearly the same as the one between the red energy levels of group 1 and 3.
For all the energy levels that shift in pairs in Fig. \ref{d20}(d) the splitting is due to a difference of the expectation value of the electron-electron interaction energy
calculated for a symmetric and antisymmetric spatial wave function, i.e.  to the exchange integral \cite{burkard}.

\section{Asymmetric DQD}

For the symmetric system of quantum dots we described above, the splitting of the energy levels is uniquely due to the interdot tunnel coupling. In an experimental situation a deviation from the ideal symmetry is inevitable.  
For a generalization we consider the case of an asymmetric quantum dots. The asymmetry is introduced with confinement potential that includes an in-plane electric field 
\begin{flalign}
&V_{QD} (x,y)=\\ \nonumber
&-V\exp(-((|x|-d)^2+y^2)/R^2)\cdot \left (1-\phi\frac{x}{d}  \right ),
 \label{qdpot1}
\end{flalign}
with $V$ and $R$ kept unchanged. The value of the electric field is controlled by a dimensionless $\phi$ parameter, and $d$ in the 
denominator of the expression in brackets keeps the potential of the left and right dot in a fixed offset as the interdot distance $2d$ is varied. 
The in-plane field in an experimental situation corresponds to a bias between the dots that is introduced to induce the flow of the current. 

\begin{figure}[h!]

\begin{tabular}{c c}
(a)\hspace{-0.3cm} \includegraphics[scale=0.23]{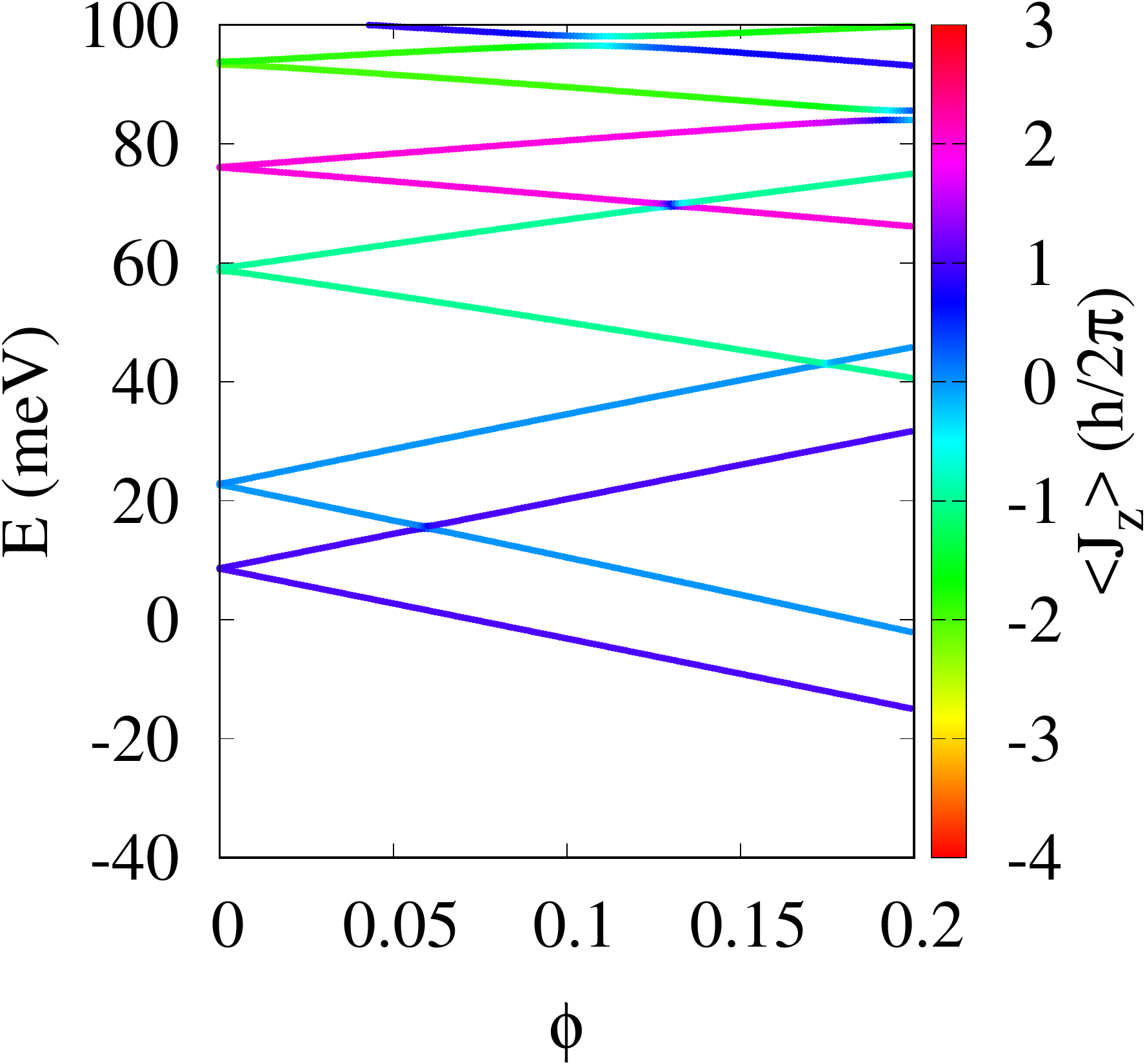} & (b)\hspace{-0.3cm}\includegraphics[scale=0.23]{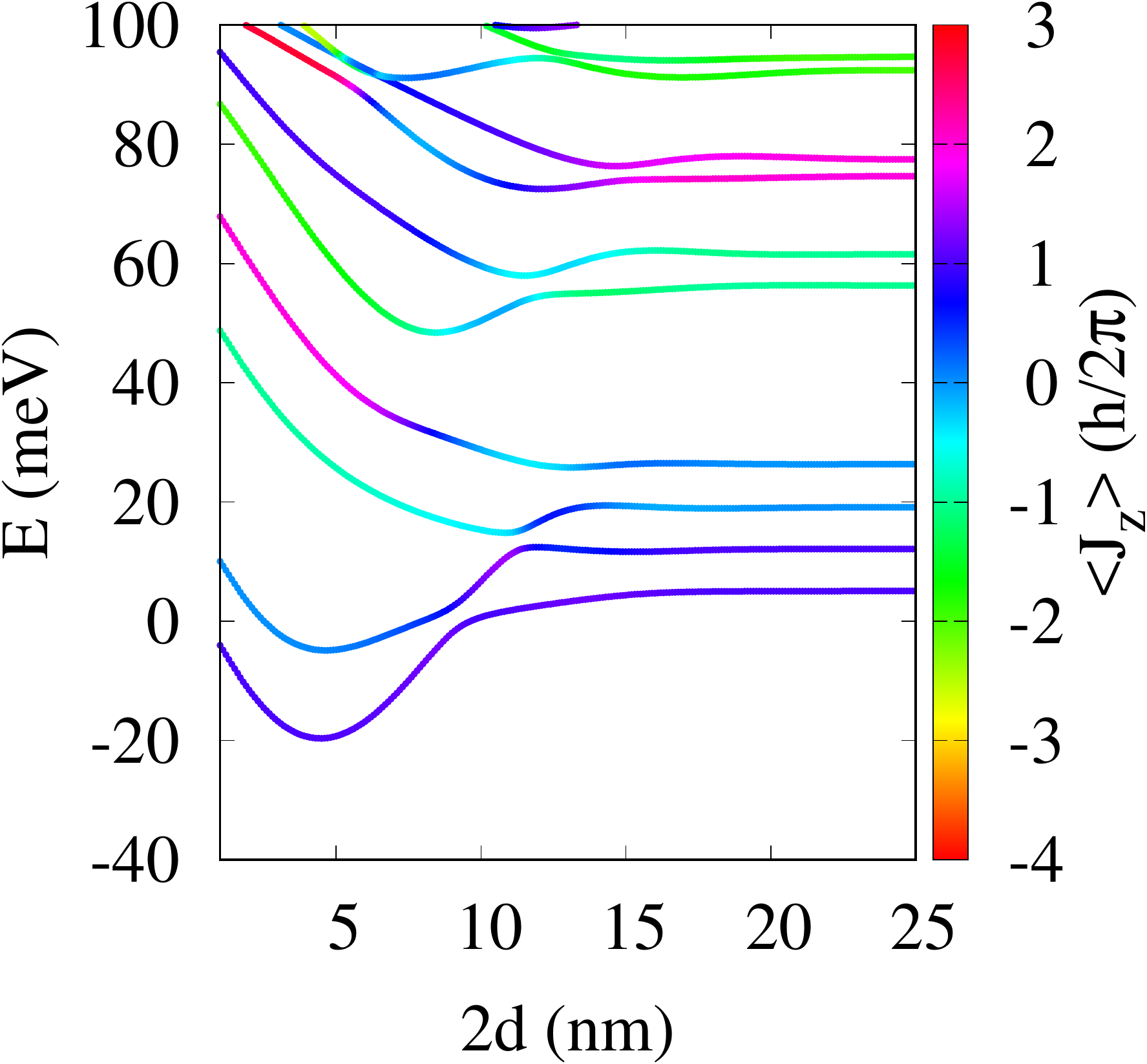} \\
\end{tabular}
\caption{ Single electron spectrum as a function of: (a) the assymetry parameter $\phi$ for the interdot distance $2d=25$ nm, (b) interdot distance for $\phi=0.03$. Results obtained by the low-energy 
continuum approximation. The color of the lines indicates the average value of the total angular momentum operator.} \label{ases}
 
\end{figure}

The single-electron spectrum presented in  Fig. \ref{ases}(a) indicates that the in-plane electric field splits the degeneracy of the energy levels at $B=0$.  For large interdot distance [Fig. \ref{ases}(b)] the single-electron states are localized either in the left or the right quantum dots. In presence of the in-plane field the angular momentum is no longer quantized, but the quantum mechanical expectation values are preserved in the weak coupling limit with respect to the symmetric case
[cf. Fig. \ref{ases}(b) and Fig. \ref{stany1}(c)]. Moreover, the parity symmetry of the external potential is broken by the in-plane electric field. The Hamiltonian eigenstates of the  $\hat{U}_P$  operator, which opens avoided crossing between the corresponding energy levels as the interdot distance is varied [Fig. \ref{ases}(b)]. Nevertheless, for a small interdot distance the tunnel coupling prevails over the interdot asymmetry and the results for both the energy levels and the angular momenta are similar to the ideal case [cf. Fig. \ref{ases}(b) and Fig. \ref{stany1}(c)].

\begin{figure}
\includegraphics[scale=0.25]{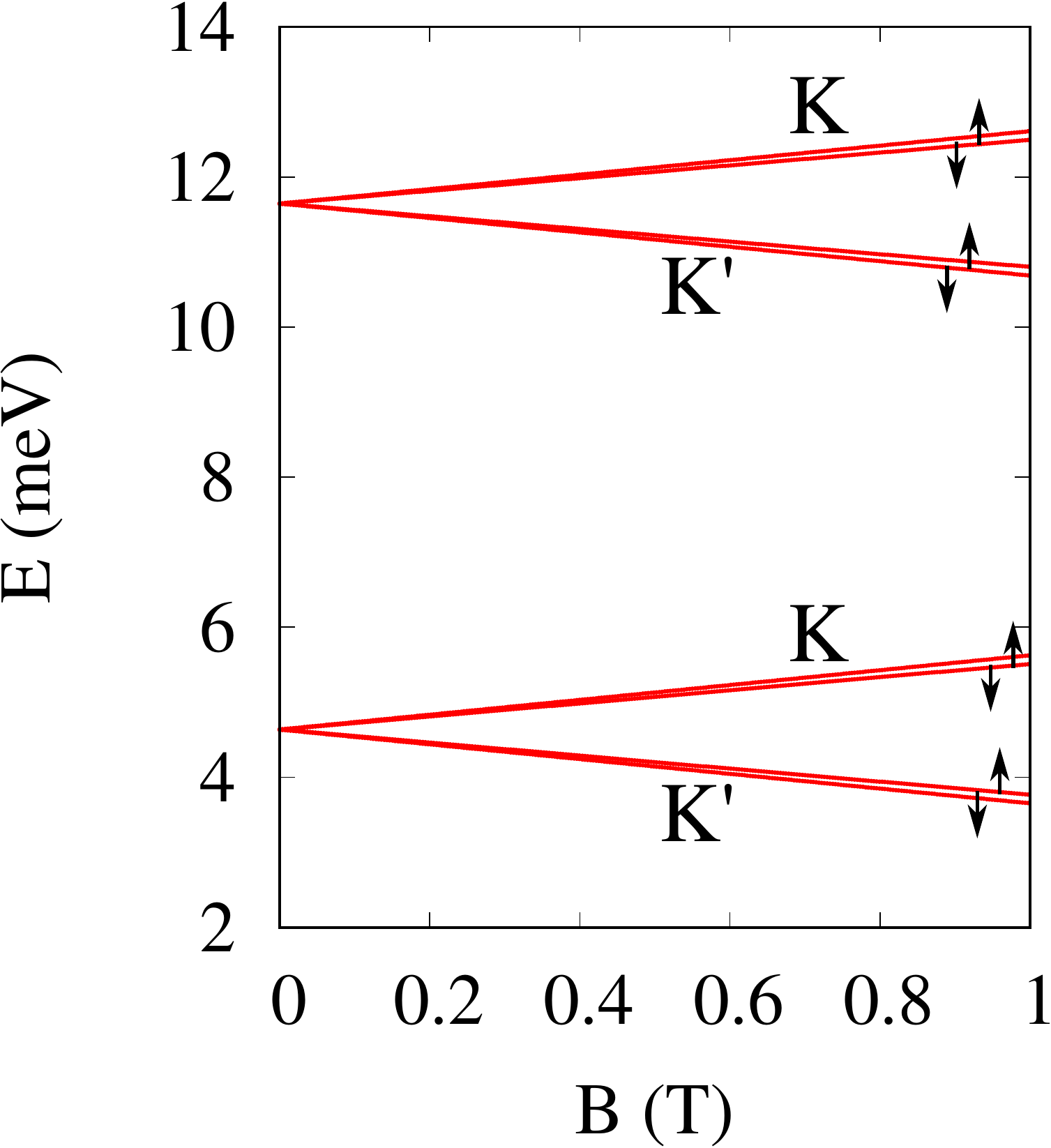}
\caption{The same as Fig. \ref{inB} (b) but for the assymetric quantum dots with $\phi=0.03$.}\label{wbas}
\end{figure}

Figure \ref{wbas} shows the spectrum in the external magnetic field. The asymmetry of the confinement potential increases the 
energy spacing between the degenerate doublets at $B=0$ displacing the crossing of the $K$ and $K'$ valley energy levels
to higher values of the magnetic field (compare with Fig. \ref{inB}). 

For the low-energy states of the two-electron system at $d\simeq 20$ nm the electrons occupy separate dots. The energy level in one of the dots is shifted up and the other down, so no change to the ground-state energy level structure is observed unless the offset of the quantum dot potentials exceeds 60 meV, where the states with the electrons in the same dot occupy the same dot with the energy spectrum of the form presented in Fig. 6(b).

\section{Summary and conclusions}

We have considered formation of extended orbitals in bilayer graphene quantum dots using an atomistic tight-binding and the continuum approach. 
The various angular momentum in each of the four sublattices for the total angular momentum eigenstates of a circular quantum dot evolve into various spatial parities
for each of the wave function components for the double quantum dot system. The symmetry amounts in mixed bonding and antibonding character of wave function
on separate sublattices and a complex dependence of the energy spectrum on the interdot distance. Both the flake large enough to be considered infinite for the dot-localized
states and a smaller flake for which the valley mixing effects of the boundary could be described.

For the electron pair we used the configuration interaction method based on the atomistic single-electron wave functions. We have reproduced the limit result of the spectrum for a single quantum dot 
when the DQDs are close to one another and  identified the effects of the interdot tunnel coupling between the dots for the more realistic weaker interdot tunnel coupling. 
The effects include -- the splitting of energy levels in the high energy part of the spectrum where the electron pair forms bonding and antibonding two-electron orbitals,
and near the ground-state -- the exchange energy that splits the symmetric and antisymmetric pairs of energy levels that shift parallel in the external magnetic field.
\section*{Acknowledgments}
This work was supported by the National Science Centre
according to decision DEC-2013/11/B/ST3/03837 and by the Flemish Science Foundation (FWO-VL).  Calculations were performed
in the PL-Grid Infrastructure.

\bibliographystyle{unsrt}

\end{document}